\documentclass[a4paper,11pt,twoside]{article}
\usepackage{graphicx}
\usepackage{pdflscape}
\usepackage[square,sort,comma,numbers]{natbib}
\usepackage{authblk}

\renewcommand{\rmdefault}{cmss} %helvetica
\fontfamily{\rmdefault}
\selectfont

\begin{document}

\title{INPOP new release: INPOP13c}

 \author[1,2]{A. Fienga}
 \author[1]{H. Manche}
 \author[1]{J. Laskar}
 \author[1]{M. Gastineau}
 \author[2]{A. Verma}
 %\and A. Verma\inst{2}
 %\and P. Kuchynka\inst{1,3}
\affil[1]{Astronomie et Syst\`emes Dynamiques, IMCCE-CNRS UMR8028, 77 Av. Denfert-Rochereau, 75014 Paris, France}
\affil[2]{AstroG\'eo, G\'eoazur-CNRS UMR7329, Observatoire de la C™te d'Azur, 250 avenue A. Einstein, 06560 Valbonne, France}
%\email{agnes.fienga@obs-besancon.fr}
%\renewcommand\Authands{ and }

% \offprints{A. Fienga}
%\institute{Astronomie et Syst\`emes Dynamiques,
% IMCCE-CNRS UMR8028,
% 77 Av. Denfert-Rochereau, 75014 Paris, France
% \and
% Observatoire de Besan\c con, CNRS UMR6213,
% 41bis Av. de l'Observatoire, 25000 Besan\c con, France
% }

 %\offprints{A. Fienga, agnes.fienga@obs-besancon.fr}

 \date{}

% \titlerunning{ INPOP10}
% \authorrunning{Fienga et al}

%  \abstract{

% \keywords{celestial mechanics - ephemerides}
% }
  \maketitle

%%-----------------------------------------------------------------
%%        The abstract
%%
%%  Warning!  within the abstract:
%%  - do not use macros.
%%  - do not use commands like: \cite, \citet, \citep ... etc.

\begin{abstract}
Based on the use of MESSENGER radiotracking data in the construction of new Mercury ephemerides (\cite{2014A&A...561A.115V}) a new planetary ephemerides INPOP13c was built including Mercury improvements but also improvements on the Mars orbit and on the tie of INPOP planetary ephemerides to ICRF in general.
\end{abstract}

%% Insert the keywords (to appear in the ADS indexing)
%% Keywords must be separated by a comma
%\begin{keywords}
%Planetary ephemerides, numerical integration, space missions, tests of general relativity, asteroid masses
%\end{keywords}

%%-----------------------------------------------------------------

\section{INPOP13c}
%%---------------------

The MESSENGER mission was the first mission dedicated to the study on Mercury. The spacecraft orbits the smallest and the Sun closest planet  of the solars system since 2011.
(\cite{2014A&A...561A.115V}) described the methods and procedures used for the analysis of the MESSENGER Doppler and range data included in the construction of the Mercury improved ephemerides, INPOP13a.

INPOP13c is an upgraded version of INPOP13a, fitted to LLR observations, and including new observations of Mars and Venus deduced from MEX, Mars Odyssey and VEX tracking (\cite{M2012}, \cite{M2013} , \cite{Marty13}).  The Pluto orbit  was also improved with a better iterative procedure.
Tables \ref{omctab0} and \ref{omctab1} resume the data samples and the obtained residuals with INPOP13c and INPOP10e common to the two ephemerides when Table \ref{omcsupp} exhibits the residuals obtained for the data samples added since INPOP10e.
Thanks to this supplementary material, a better extrapolation capability of the Mars ephemerides appears clearly in Tables \ref{omcsupp}, \ref{comparomc1} and \ref{comparomc2} as well as better consistencies between DE and INPOP ephemerides (see for example Table \ref{xyz0}). The section \ref{uncert} presents comparisons between INPOP13c, INPOP10e (\cite{INPOP10e}) and DE430 (\cite{2014IPNPR.196C...1F}).

Adjustment of the gravitational mass of the sun was performed as recommended by the IAU resolution B2 as well as the sun oblateness (J$_{2}$), the ratio between the mass of the earth and the mass of the moon (EMRAT) and the mass of the Earth-Moon barycenter. Estimated values are presented on Table \ref{paramfita}.

Masses of planets as well as procedures of estimations of the asteroid masses perturbing the inner planet orbits are the same as in INPOP10e and INPOP13a. Comparisons between asteroid masses obtained with INPOP13c and values gathered in \cite{CARRY2012} are given and discussed in section \ref{sect_ast}.

\begin{table}
\caption{Values of parameters obtained in the fit of INPOP13c, INPOP10e and DE430 to observations.}
\begin{center}
\begin{tabular}{l c c c }
\hline
&  INPOP13c & INPOP10e & DE430 \\
&    $\pm$ 1$\sigma$ & $\pm$ 1$\sigma$ & $\pm$ 1$\sigma$ \\
\hline
(EMRAT-81.3000)$\times$ 10$^{-4}$ &  (5.694 $\pm$ 0.010) & (5.700 $\pm$ 0.020) & ($\pm$) \\

J$_{2}$$^{\odot}$ $\times$ 10$^{-7}$ & (2.30 $\pm$ 0.25) & (1.80 $\pm$ 0.25)  & 1.80  \\
\hline
GM$_{\odot}$ - 132712440000 [km$^{3}.$ s$^{-2}$]&  (44.487 $\pm$ 0.17) & (50.16 $\pm$ 1.3) & 40.944 \\
AU - 1.49597870700 $\times$ 10$^{11}$ [m] & 0.0 & 0.0 & (-0.3738 $\pm$ 3 ) \\
\hline
 [M$_{\odot}$ / M$_{\textrm{EMB}}$] - 328900 & 0.55314 $\pm$ 0.00033  &  0.55223 $\pm$ 0.004  &  0.55915 $\pm$ NC \\
 \hline
 \end{tabular}
%\hline
\end{center}
\label{paramfita}
\end{table}

%Thanks to the added solar corrections and to the improvement in the fit procedure, 152 asteroid masses have been estimated (see section \ref{secapp}).
%Comparisons with values of asteroid masses found in the literature show a good agreement with INPOP10e ones, especially for bodies impacting the earth-Mars distances with more than 5 meters over the observational period .
% Improvement in the INPOP extrapolation capabilities is also achieved with this new version.
%Comparisons to other planetary ephemerides, postfit and extrapolated residuals are discussed in section \ref{uncert}.

\section{Estimation of uncertainties}
\label{uncert}
%%-------------------------

%\subsection{Comparisons to other planetary ephemerides}

\begin{figure}
\begin{center}
Mercury\\
\includegraphics[scale=0.9]{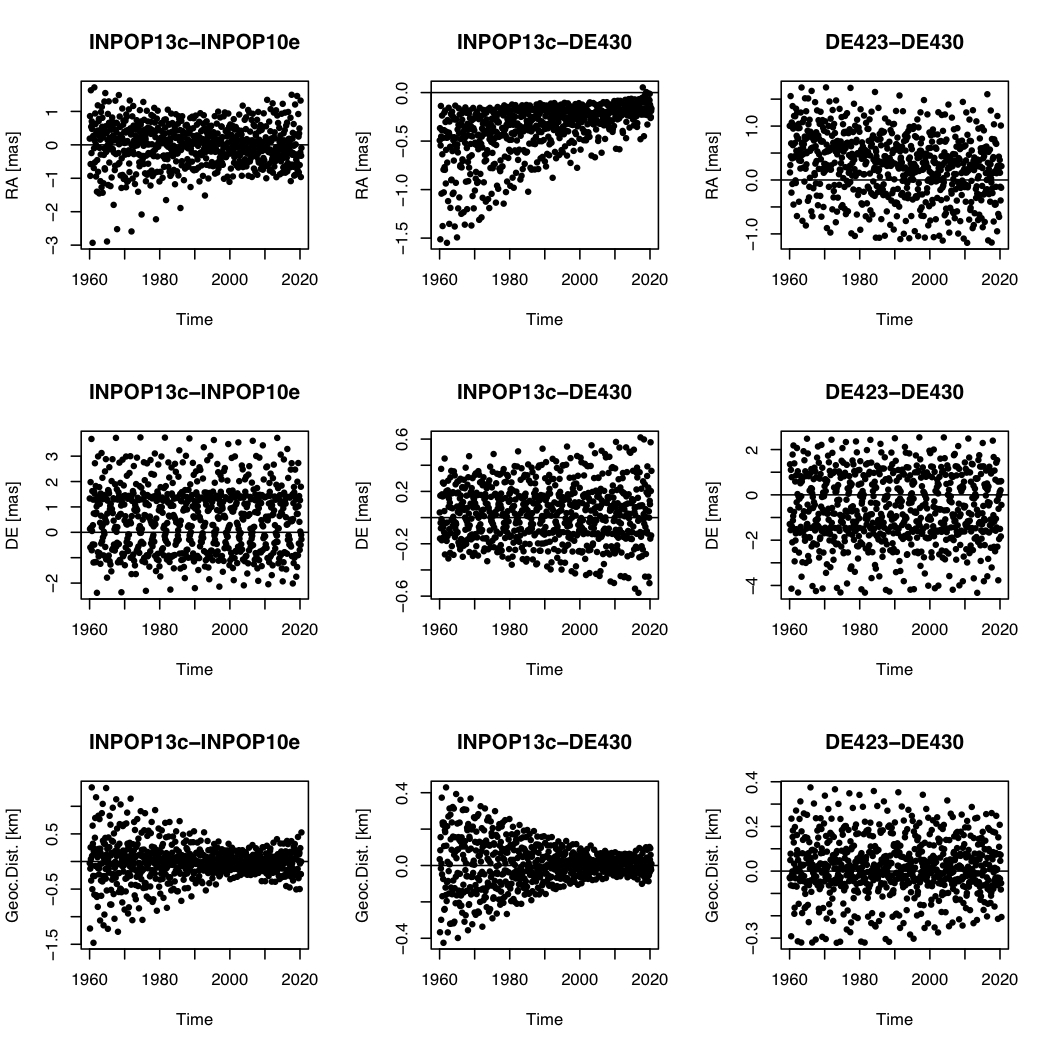}
\end{center}
\caption{Differences in $\alpha,\delta$ and geocentric distances between INPOP13c, INPOP10e, DE430 and DE423.}
\label{plot1}
\end{figure}

\begin{figure}
\begin{center}
Venus\\
\includegraphics[scale=0.9]{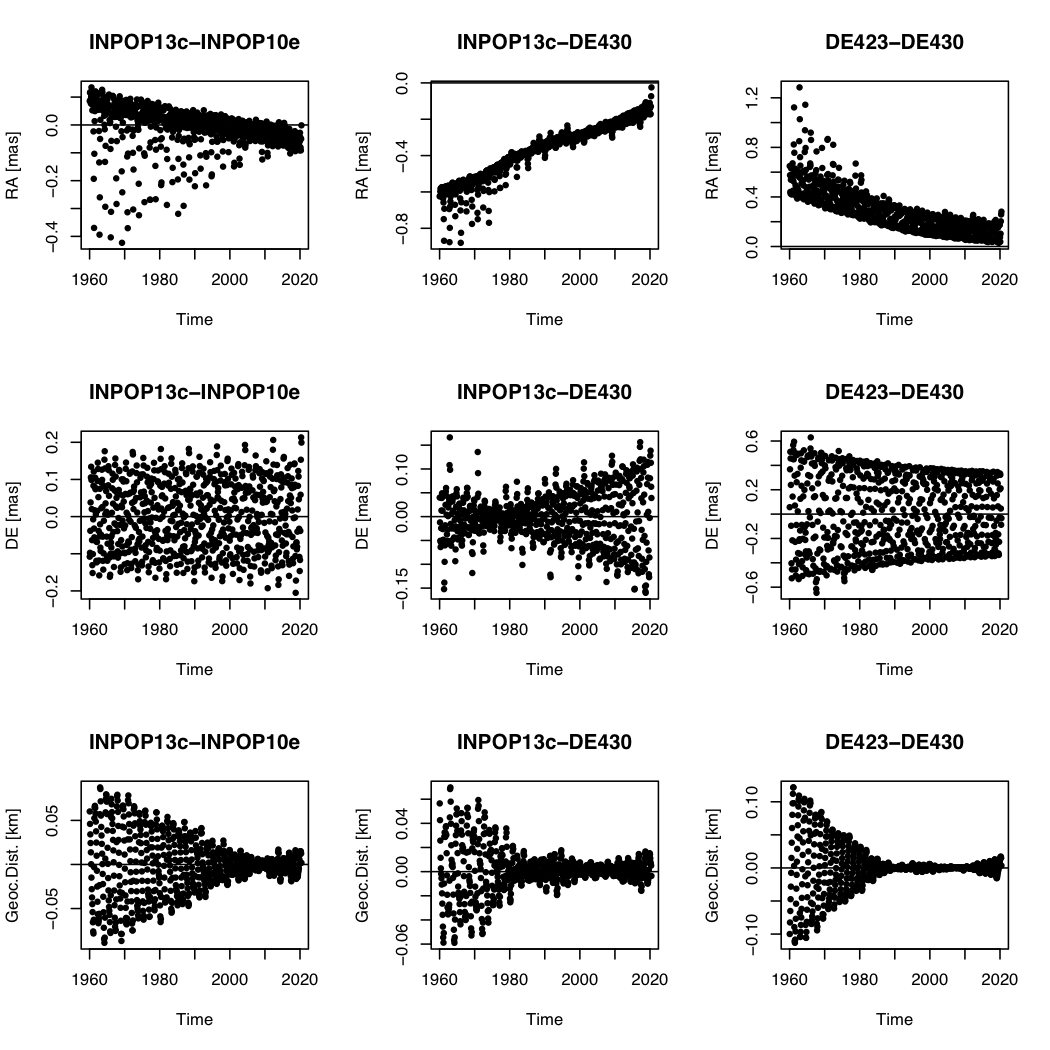}
\end{center}
\caption{Differences in $\alpha,\delta$ and geocentric distances between INPOP13c, INPOP10e, DE430 and DE423.}
\label{plot2}
\end{figure}

\begin{figure}
\begin{center}
Mars\\
\includegraphics[scale=0.9]{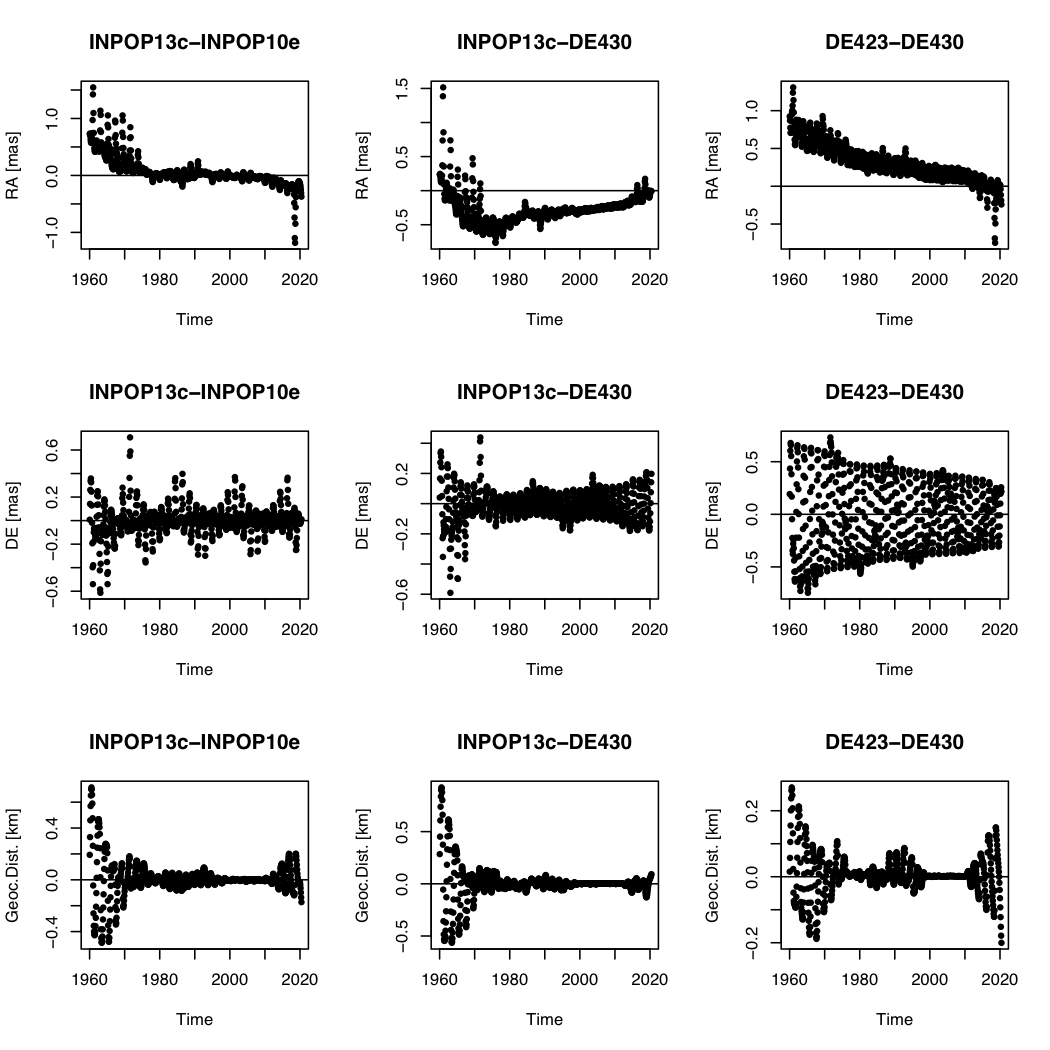}
\end{center}
\caption{Differences in $\alpha,\delta$ and geocentric distances between INPOP13c, INPOP10e, DE430 and DE423.}
\label{plot3}
\end{figure}

\begin{figure}
\begin{center}
Jupiter\\
\includegraphics[scale=0.9]{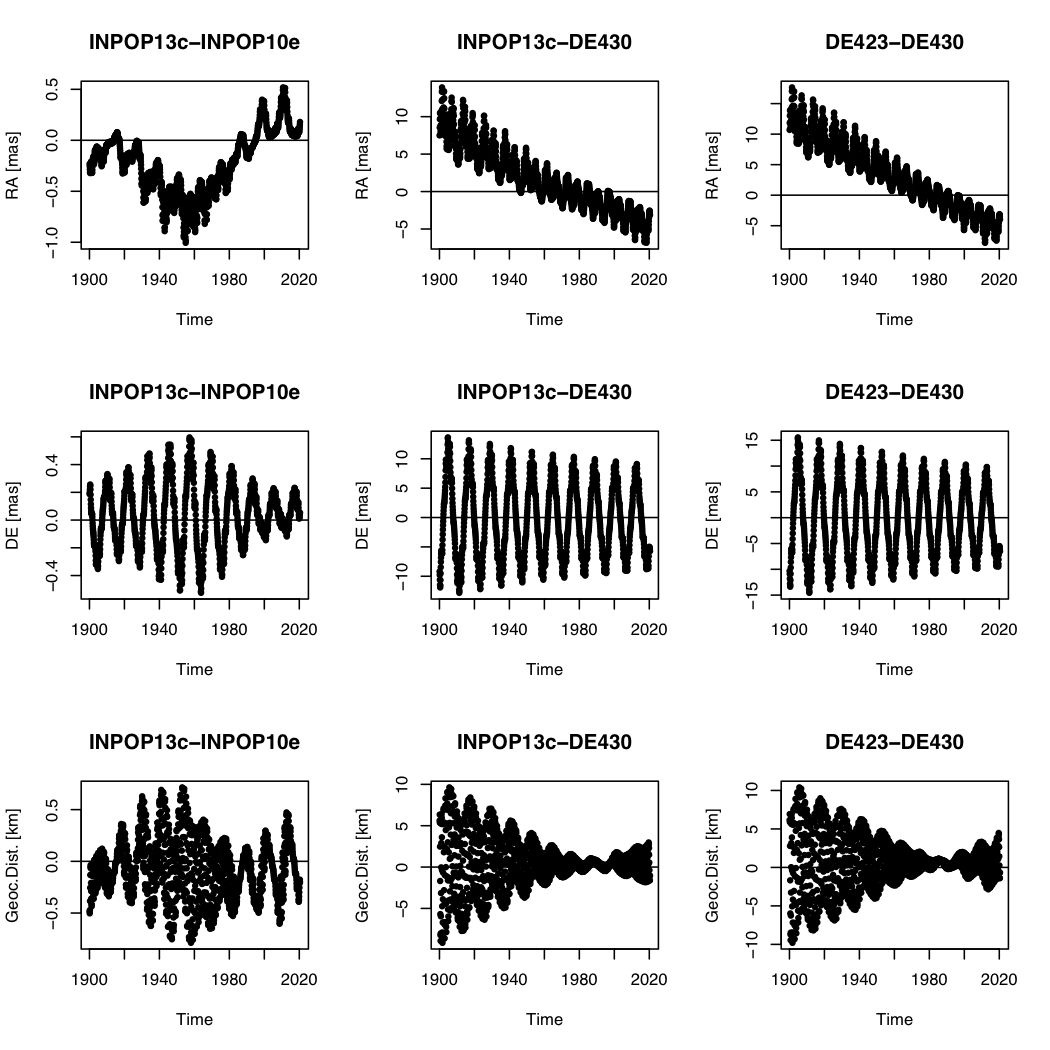}
\end{center}
\caption{Differences in $\alpha,\delta$ and geocentric distances between INPOP13c, INPOP10e, DE430 and DE423.}
\label{plot4}
\end{figure}

\begin{figure}
\begin{center}
Saturn\\
\includegraphics[scale=0.9]{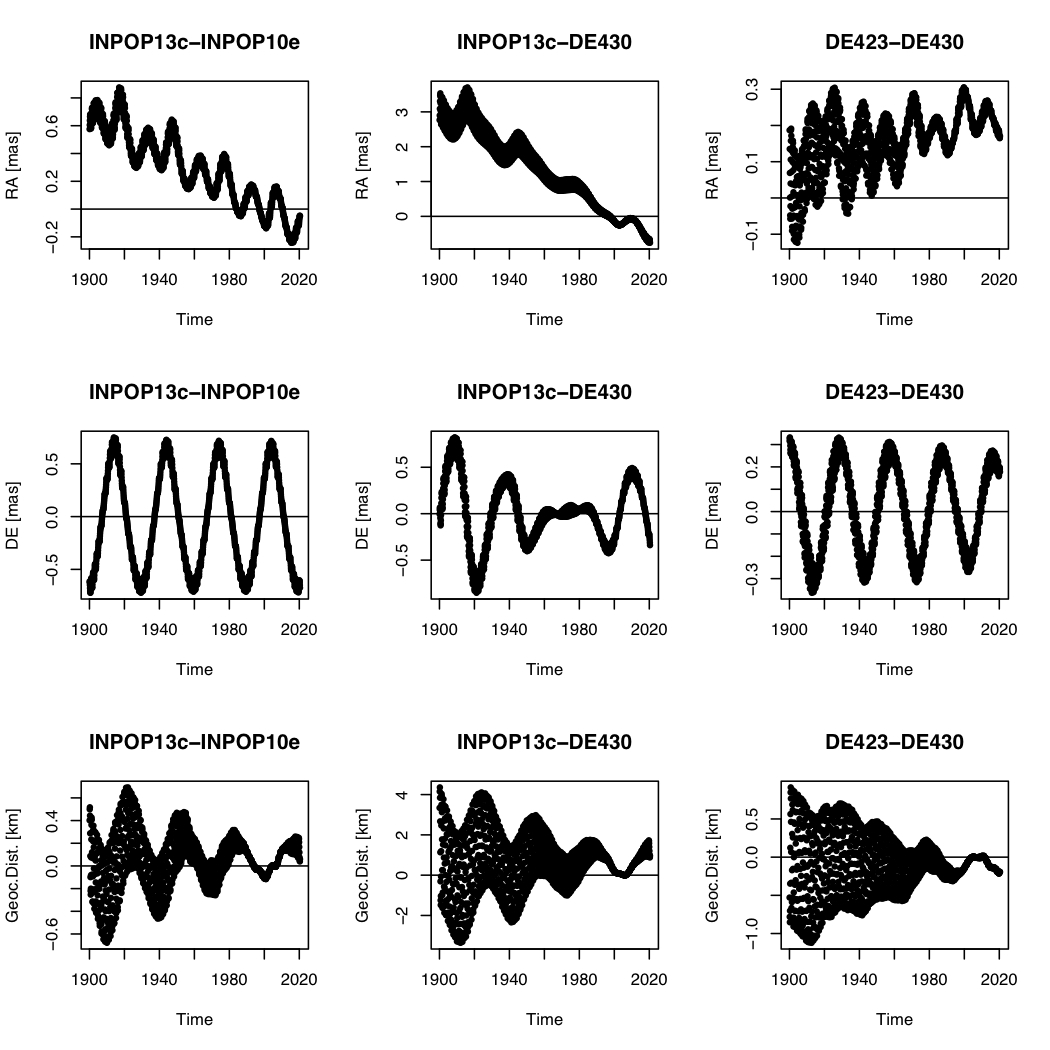}
\end{center}
\caption{Differences in $\alpha,\delta$ and geocentric distances between INPOP13c, INPOP10e, DE430 and DE423.}
\label{plot5}
\end{figure}

\begin{figure}
\begin{center}
Uranus\\
\includegraphics[scale=0.9]{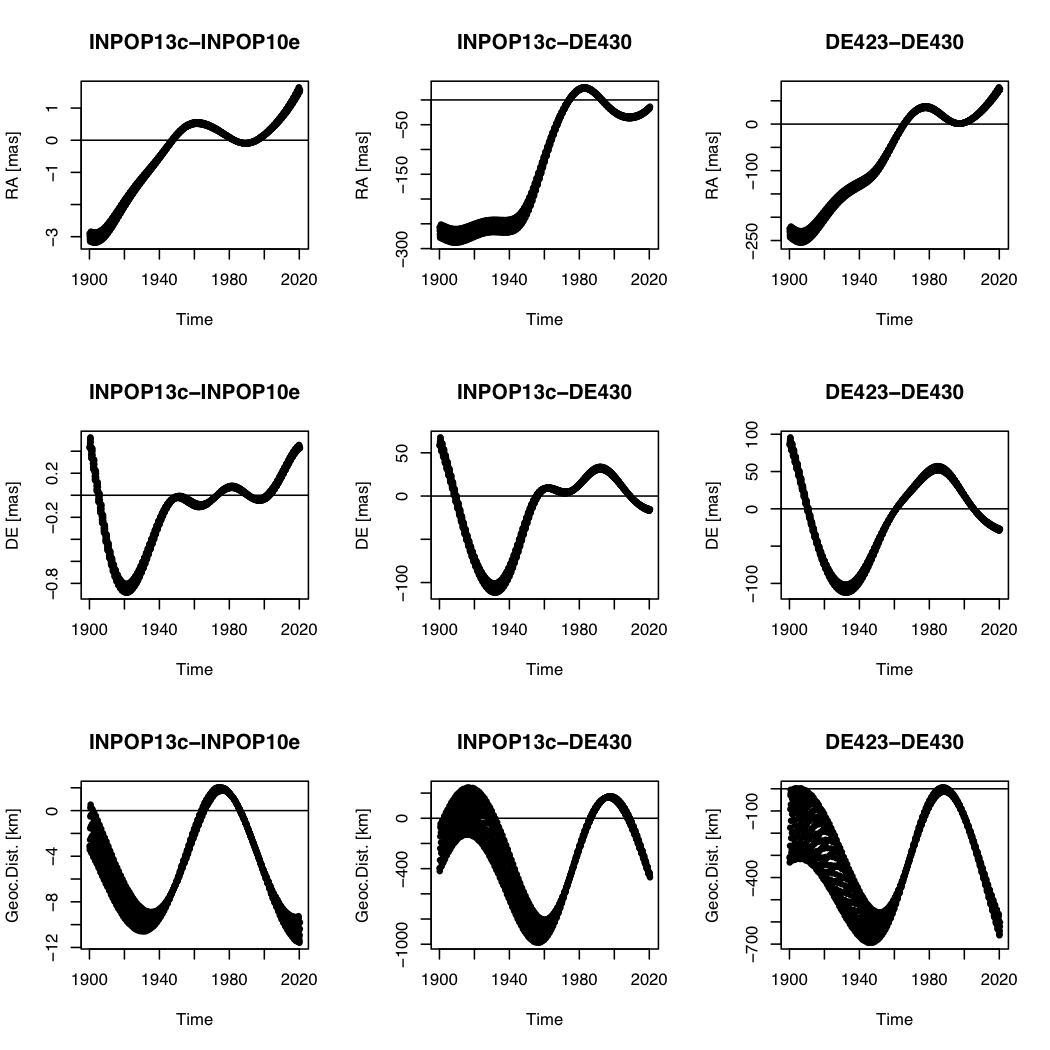}
\end{center}
\caption{Differences in $\alpha,\delta$ and geocentric distances between INPOP13c, INPOP10e, DE430 and DE423.}
\label{plot6}
\end{figure}

\begin{figure}
\begin{center}
Neptune\\
\includegraphics[scale=0.9]{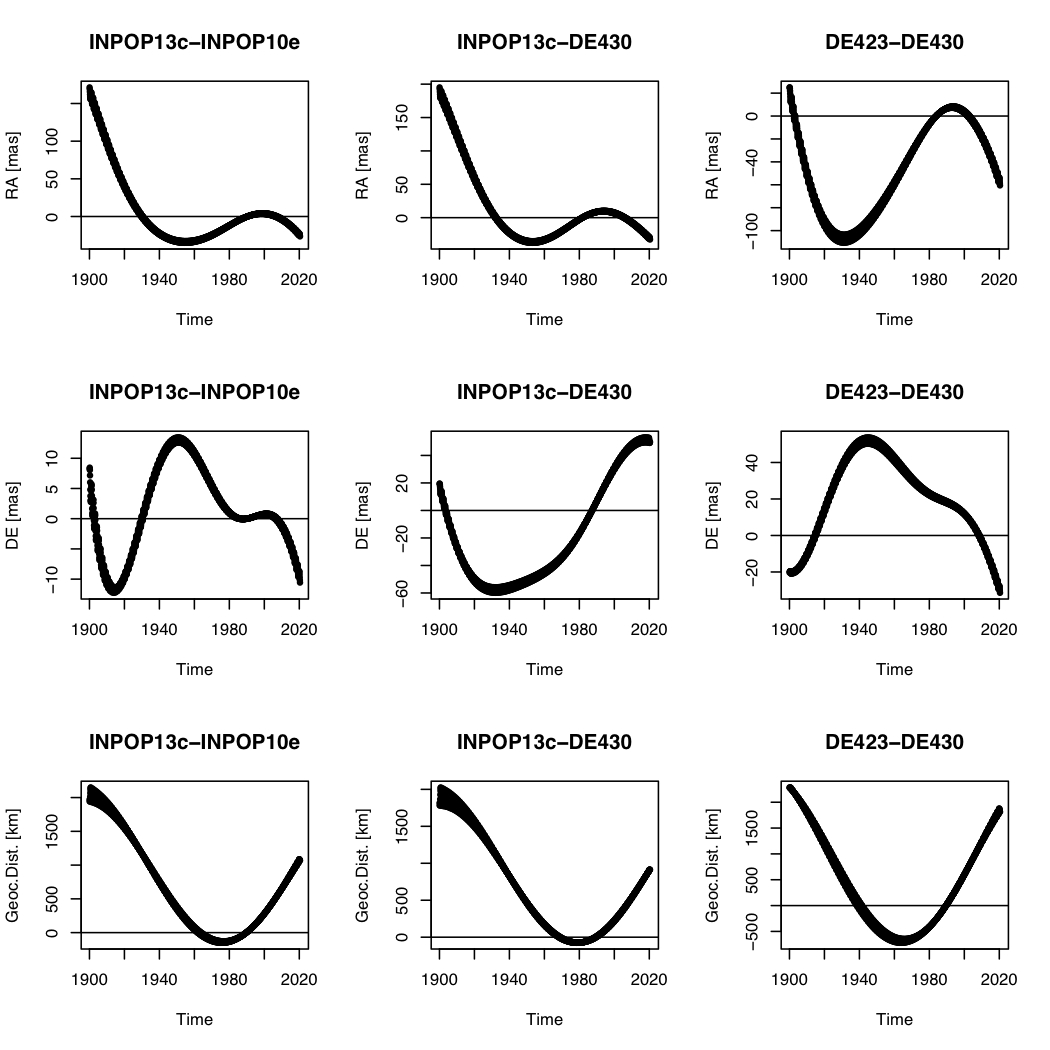}
\end{center}
\caption{Differences in $\alpha,\delta$ and geocentric distances between INPOP13c, INPOP10e, DE430 and DE423.}
\label{plot7}
\end{figure}

\begin{figure}
\begin{center}
Pluto\\
\includegraphics[scale=0.9]{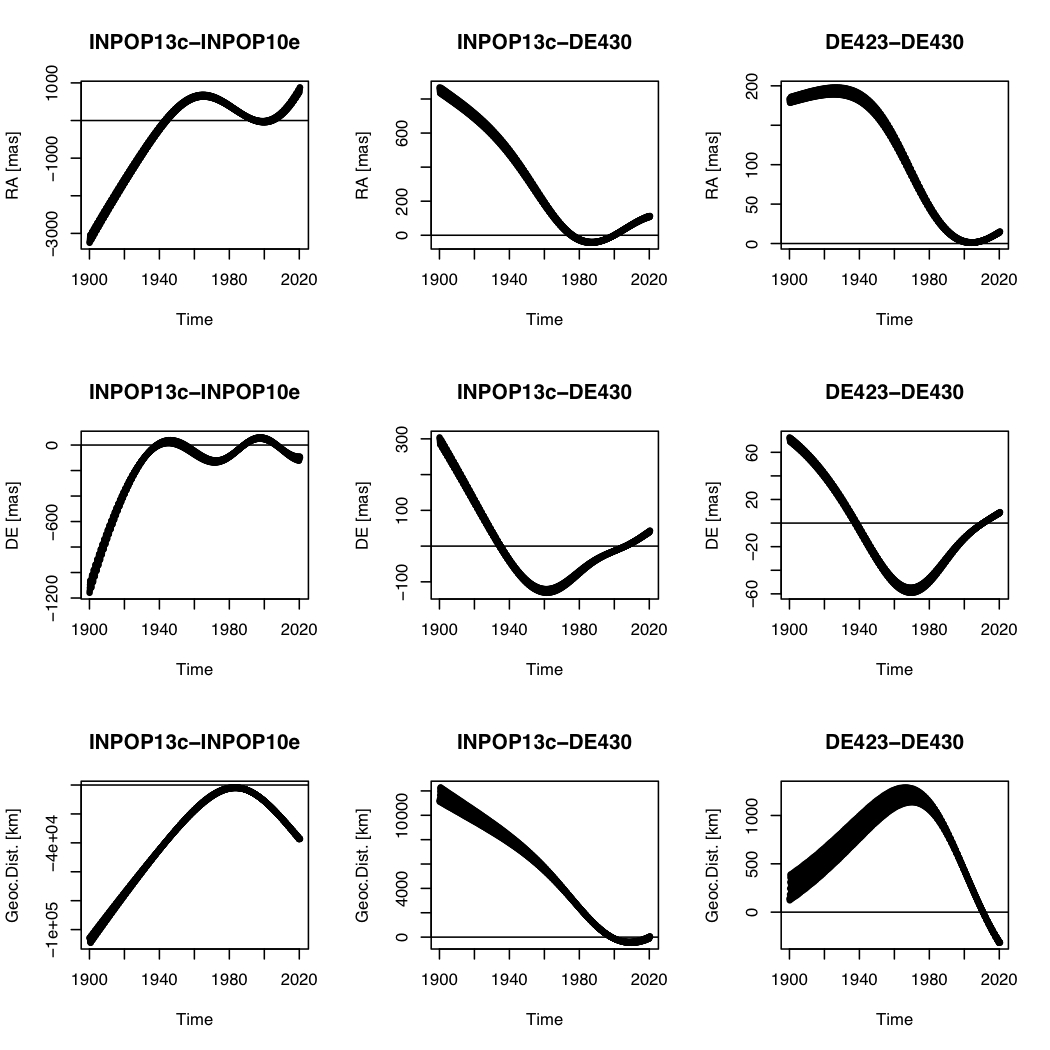}
\end{center}
\caption{Differences in $\alpha,\delta$ and geocentric distances between INPOP13c, INPOP10e, DE430 and DE423.}
\label{plot8}
\end{figure}

\begin{table}
\caption{Maximum differences between INPOP13c, INPOP10e and DE430 from 1980 to 2020 in in $\alpha,\delta$ and geocentric distances.}
\label{xyz0}
%\begin{center}
\begin{tabular}{c | c c c | c c c | c c c}
\hline
Geocentric & \multicolumn{3}{c}{INPOP13c - INPOP10e} & \multicolumn{3}{c}{INPOP13c - DE430} & \multicolumn{3}{c}{INPOP10e - DE423}\\
Differences & \multicolumn{3}{c}{1980-2020} & \multicolumn{3}{c}{1980-2020}& \multicolumn{3}{c}{1980-2020} \\
& $\alpha$ & $\delta$ & $\rho$ & $\alpha$ & $\delta$ & $\rho$& $\alpha$ & $\delta$ & $\rho$  \\
& mas & mas & km & mas & mas & km& mas & mas & km \\
Mercury & 1.4 & 3.8 & 0.51 & 0.69 & 0.61 & 0.12   & 1.58 & 1.7 & 0.65\\
Venus & 0.17 & 0.22 & 0.023 & 0.33 & 0.16 & 0.017 & 0.85 & 0.42 & 0.045\\
%moon & & & & & & & &\\
Mars & 1.19 & 0.38 & 0.204 & 0.35 & 0.21 & 0.129 & 2.1 & 0.62 & 0.47\\
Jupiter & 0.52 & 0.23 & 0.60 & 6.85 & 9.55 & 2.98  & 0.81 & 0.74 & 1.11\\
Saturn & 0.24 & 0.72 & 0.25 & 0.75 & 0.49 & 1.72 & 0.82 & 0.53 & 1.82\\
Uranus & 1.64 & 0.46 & 11.52 & 37.0 & 24.1 & 442.47 & 98.1 & 38.9 & 359.73\\
Neptune & 24.4 & 9.65 & 1081 & 31.0 & 53.1 & 904.1  & 51.0 & 91.3 & 2054.8\\
Pluton & 800.0 & 122.3 & 37384.1& 112.0& 39.8& 465.4& 703.2& 152.7& 37578.6\\
\hline
\end{tabular}
%\end{center}
\end{table}

\begin{table}
\caption{Maximum differences between INPOP13c and other planetary ephemerides from 1980 to 2020 in cartesian coordinates of the earth in the BCRS.}
\label{xyz}
\begin{center}
\begin{tabular}{c | c c}
\hline
Earth Barycentric &  XYZ & VxVyVz\\
Differences &   & \\
& km  & mm.s$^{-1}$ \\
\hline
INPOP13c - INPOP10e & 0.104 & 0.0177\\
INPOP13c - DE430 & 0.3763 & 0.0467 \\
INPOP10e-DE423 & 0.84 &  0.113\\
%INPOP10e-INPOP10d & 0.34  &  0.050 \\
\hline
\end{tabular}
\end{center}
\end{table}

Figures \ref{plot1} to \ref{plot8} present the differences in right ascension, declination and geocentric distances between INPOP13c, INPOP10e, DE430 and DE423 (\cite{Folkner2010}) over 120 years. The comparisons between INPOP13c and INPOP10e show the improvement of INPOP13c due to the addition of new data samples. DE430 was fitted on about the same sample as INPOP13c except for Saturn Cassini normal points which were reprocessed (for more details see \cite{2014IPNPR.196C...1F}). The comparisons between INPOP13c and DE430 give then an estimation of the ephemerides uncertainties induced by the use of different procedures of data analysis and of weighting schema. The differences DE430-DE423 are also plotted.
The Tables \ref{xyz0} and \ref{xyz} give the maximum differences obtained by comparisons between the same ephemerides on a 1980 to 2020 interval.

For the inner planets the differences pictured on Figures \ref{plot1},\ref{plot2} and \ref{plot3} show a good consistency between DE430, INPOP13c and INPOP10e. \\
For Mercury, the impact of the MESSENGER range bias used for INPOP13c and DE430 construction is clear on the post-2010 period. One can notice an increase of the differences in geocentric distances between INPOP13c and INPOP10e with the time. This increase does not appear in the DE423-DE430 differences. This fact could be explained by the lack of accurate observations before 2008 and the launch of MESSENGER. It is also noticed in the comparisons of the postfit residuals of Mariner data set in the Table \ref{omctab0}.
The Table \ref{omcsupp} illustrates the improvement brought by the use of the MESSENGER range bias in the construction of INPOP13c by comparing INPOP13c postfit residuals and residuals obtained with INPOP10e.
The one-order of magnitude improvement is consistent with INPOP13a (\cite{2014A&A...561A.115V}) and is also illustrated on the Figures of the Table \ref{comparomc2}.

For Venus, one can note that the differences between DE423 and DE430 are one-order of magnitude larger than those between INPOP13c, INPOP10e and DE430 demonstrating an improvement of the accuracy of Venus orbit, especially through the computation of the geocentric distances. A common increase of the one-way range residuals of a factor 3 for INPOP13c and INPOP10 after 2010 shown on Table \ref{omcsupp} and on plots of the Table \ref{comparomc2} illustrates a confirmed degradation of the VEX transpondeur  with time.

For Mars, as one can see on Figure \ref{plot3} and on Table \ref{xyz0} the differences from one ephemeris to another one are quite equivalent for angles and geocentric distances. This illustrates the intensive works done in the past few years for improving the extrapolation capabilities of the ephemerides for the highly perturbed Mars orbit. The figures of the Tables \ref{comparomc1} and  \ref{comparomc2} also show the improvement of the smoothness of the postfit residuals obtained with INPOP13c.

In terms of reference frame, the VLBI observations of Mars orbiters give the strongest link to ICRF2 as explained in \cite{2014IPNPR.196C...1F}. As the INPOP13c-DE430 differences obtained for $\alpha$ and $\delta$ are below 0.5 mas and as the postfit residuals of INPOP13c VLBI Mars observations are also below 0.5 mas, the link of INPOP13c to ICRF2 is garenteed to at least 0.5 mas over the observational time interval (from 1989.13 to 2007.97) but also over the interval of ephemeris comparisons (from 1980 to 2020). Globally we can notice very small differences ($< 1$ mas) in angular geocentric quantities over 60 years for inner planets.

For the outer planets, differences are showed on Figures \ref{plot4}, \ref{plot5}, \ref{plot6} and \ref{plot7}. A summary of the maximum differences is given on Table \ref{xyz0} and postfit residuals are presented on Tables \ref{omctab0} and \ref{omctab1}.\\
The Jupiter differences show an non-negligeable discrepency between INPOP13c, INPOP10e and DE423 in one hand and DE430 in the other hand. This appears for all quantities ($\alpha$,$\delta$ and geocentic distances) reaching about 15 mas in $\delta$ for DE423-DE430 and INPOP13c-DE430 over 1 century and about 10 mas over the contemporary interval (from 1980 to 2020). On the opposite, the maximum differences obtained by comparing INPOP13c to INPOP10e or INPOP10e to DE423 (see Table \ref{xyz0}) do not present such offsets. These results could be explained by the change of weights used for the construction of the ephemerides. The uncertainties of the Jupiter modern observations were indeed not well estimated due to failures in the operational procedures (for example with the Galileo High Gain Antenna) or due to indirect measurements during flybys (\cite{2014IPNPR.196C...1F}). One can then wait to have better inputs from the future Junon mission which is supposed to reach Jupiter in 2015.

Despite the use of new analysed Cassini tracking data in DE430, the differences of the Saturn orbits estimated with INPOP ephemerides and DE430 are less important than the Jupiter differences. They are also consistent with the expected accuracy of the observations used to constraint the orbit: angular differences are below 1 mas thanks to VLBI tracking of the Cassini spacecraft and differences in geocentric distances are about 1 km over the observational interval.

For Uranus and Neptune, the differences INPOP13c-DE430 and DE430-DE423 are compatible with the expected accuracy of the observations used for the construction of the ephemerides as given by the Table \ref{xyz0} maximum differences for the contemporary interval and by the postfit residuals of Table \ref{omctab1}.

\begin{figure}
\label{comparomc5}
\begin{center}\includegraphics[scale=0.45]{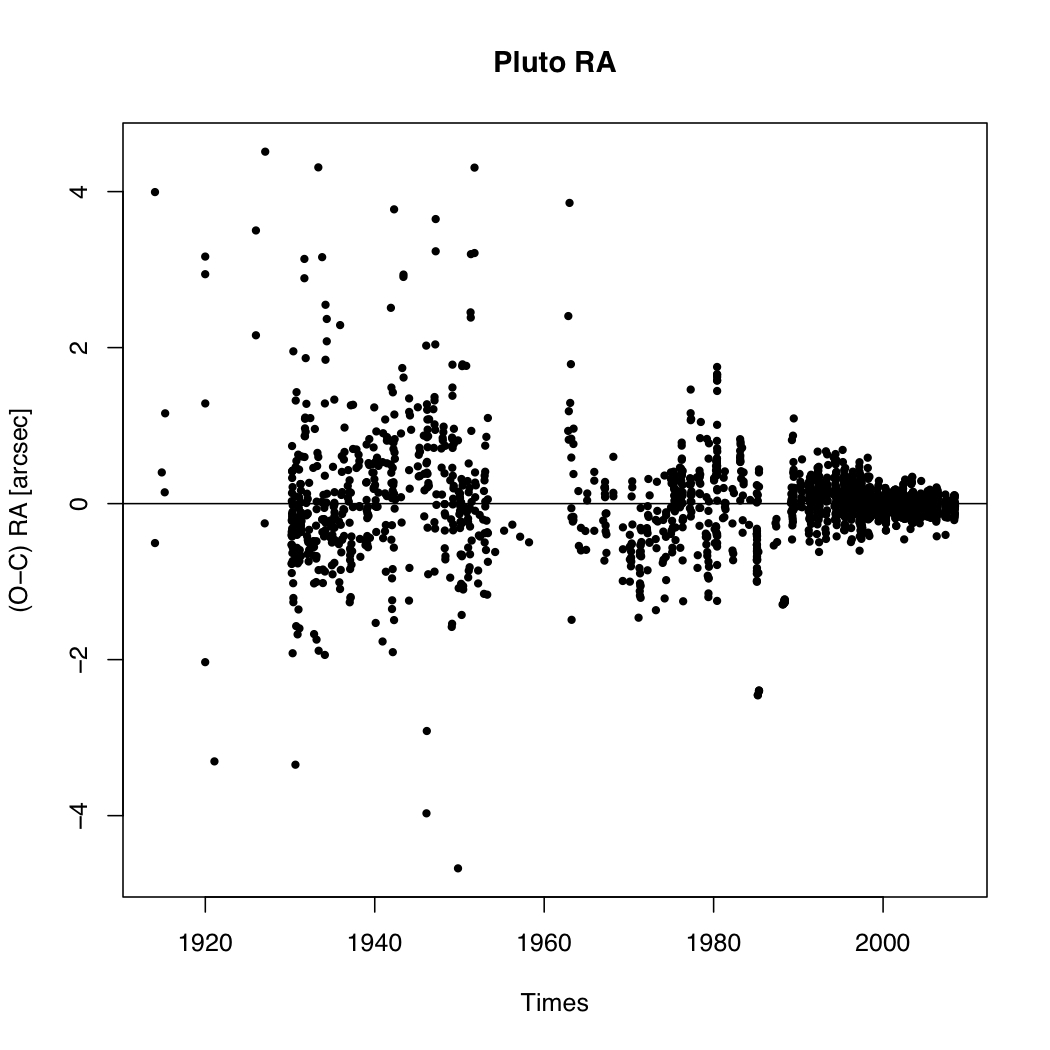}\includegraphics[scale=0.45]{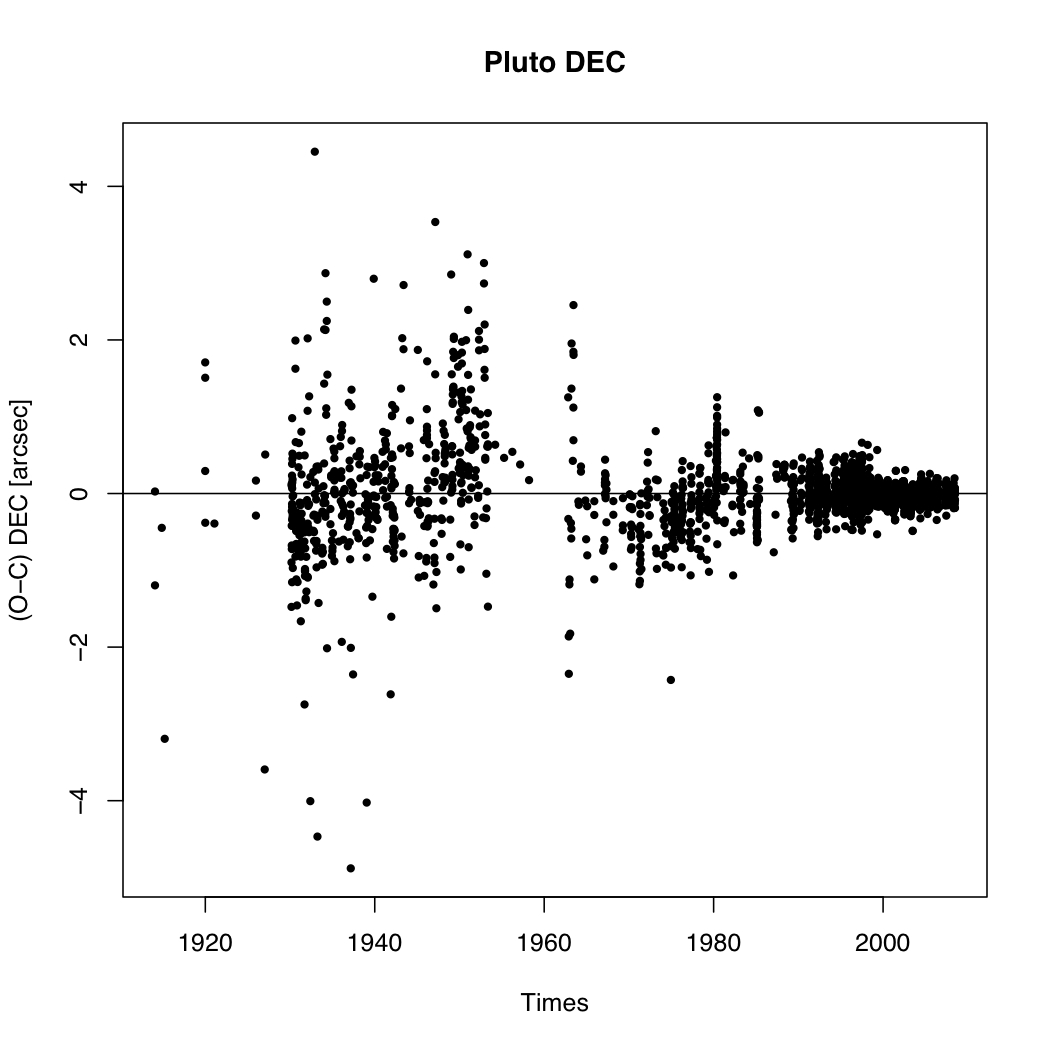}\end{center}
\caption{Pluto Postfit residuals in right ascension (RA) and  declination (DE) in arcseconds obtained with INPOP13c.}
\end{figure}

For Pluto, the large differences after 1960 pictured on Figure \ref{plot8} and on Figure \ref{comparomc5}  are essentially induced by the important dispersions of photographic residuals used before the sixties for fitting the ephemerides. By the scanning of the photographic plates and by making a new analysis in using the future Gaia catalogue, one should be able to reduce drastically the noise of these data sets and then the uncertainties of the ephemerides for Pluto orbit. The arrival of the New Horizon spacecraft should help as well especially for the estimation of the size of the darwf planet's orbit. One can note however the good agreement of the ephemerides for the recent period thanks to the stellar occultation used for INPOP and DE constructions.

%\subsection{Comparisons to observations, extrapolation and link to the ICRF}
%\label{omcsec}

\begin{table}
\begin{center}
\begin{tabular}{c c}
%\hline
INPOP13c & INPOP10e \\
%Mars orbiter one-way range [m]&\\
\includegraphics[scale=0.4]{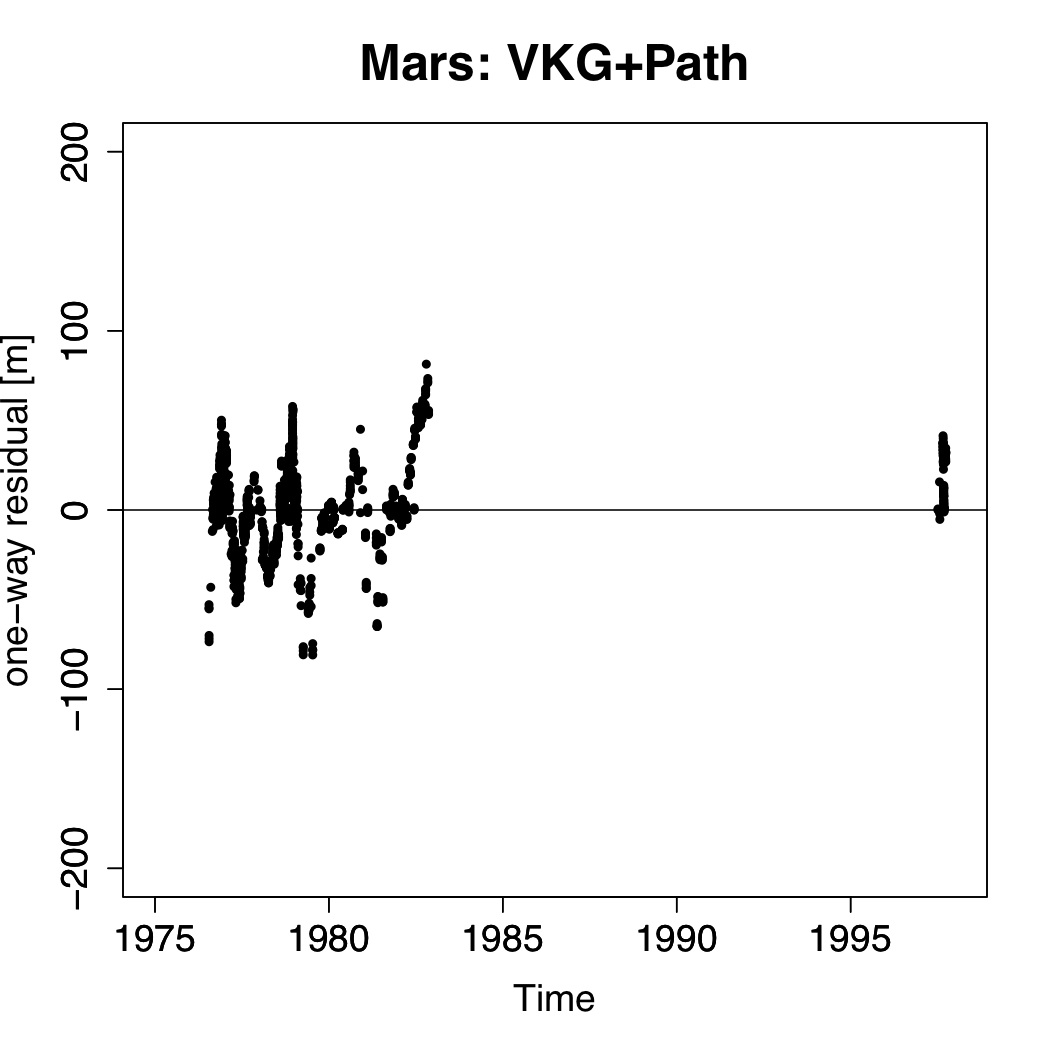}&\includegraphics[scale=0.4]{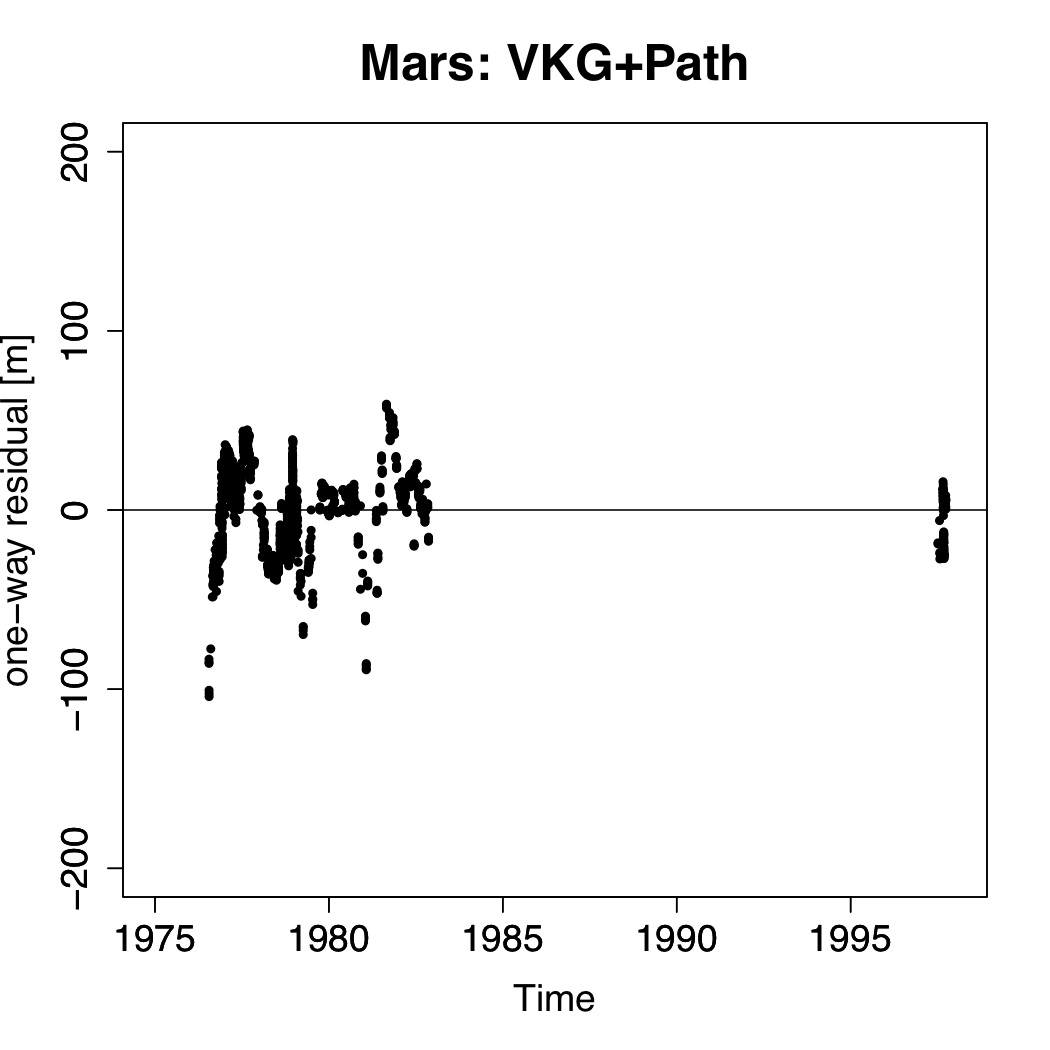}\\%&\includegraphics[scale=0.3]{marsde430.jpg}\\
\includegraphics[scale=0.4]{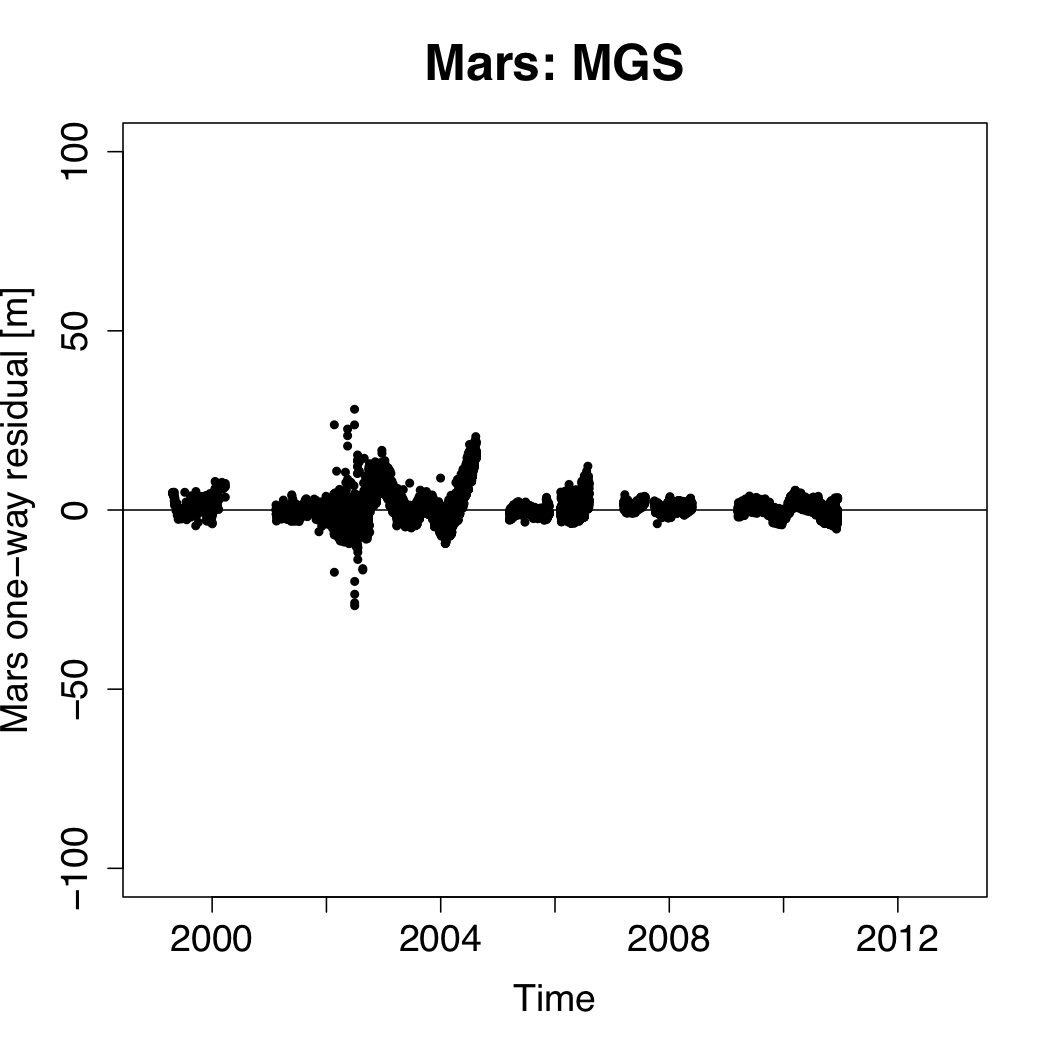}&\includegraphics[scale=0.4]{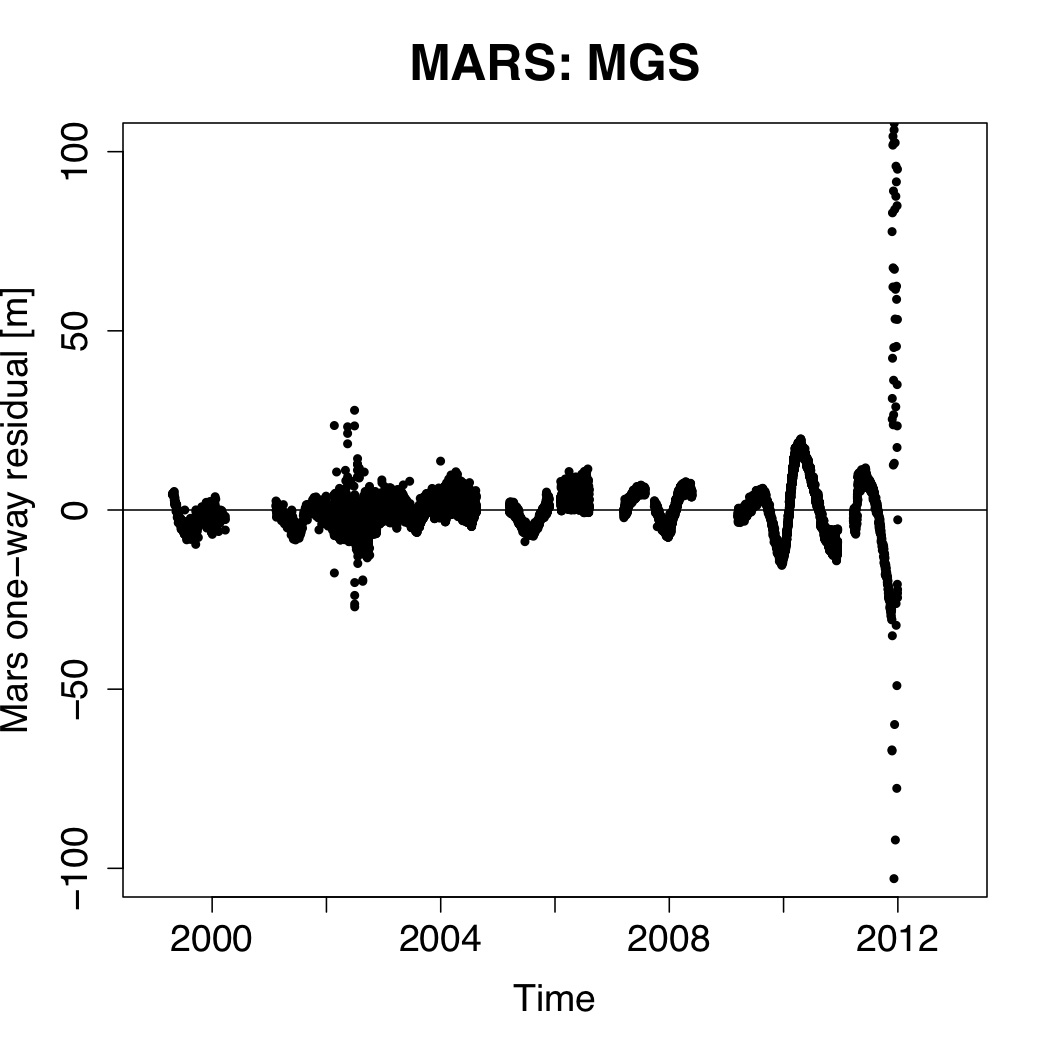}\\
\includegraphics[scale=0.4]{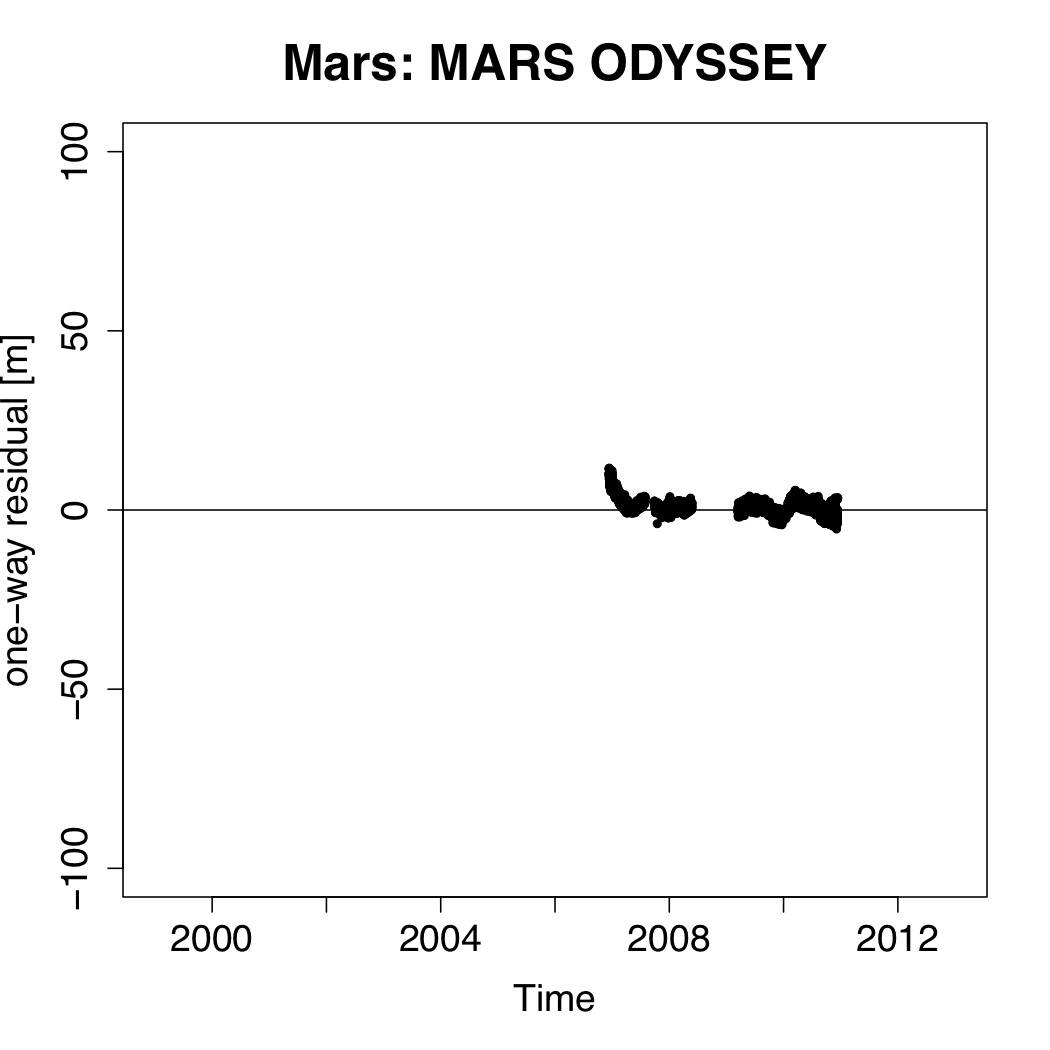}&\includegraphics[scale=0.4]{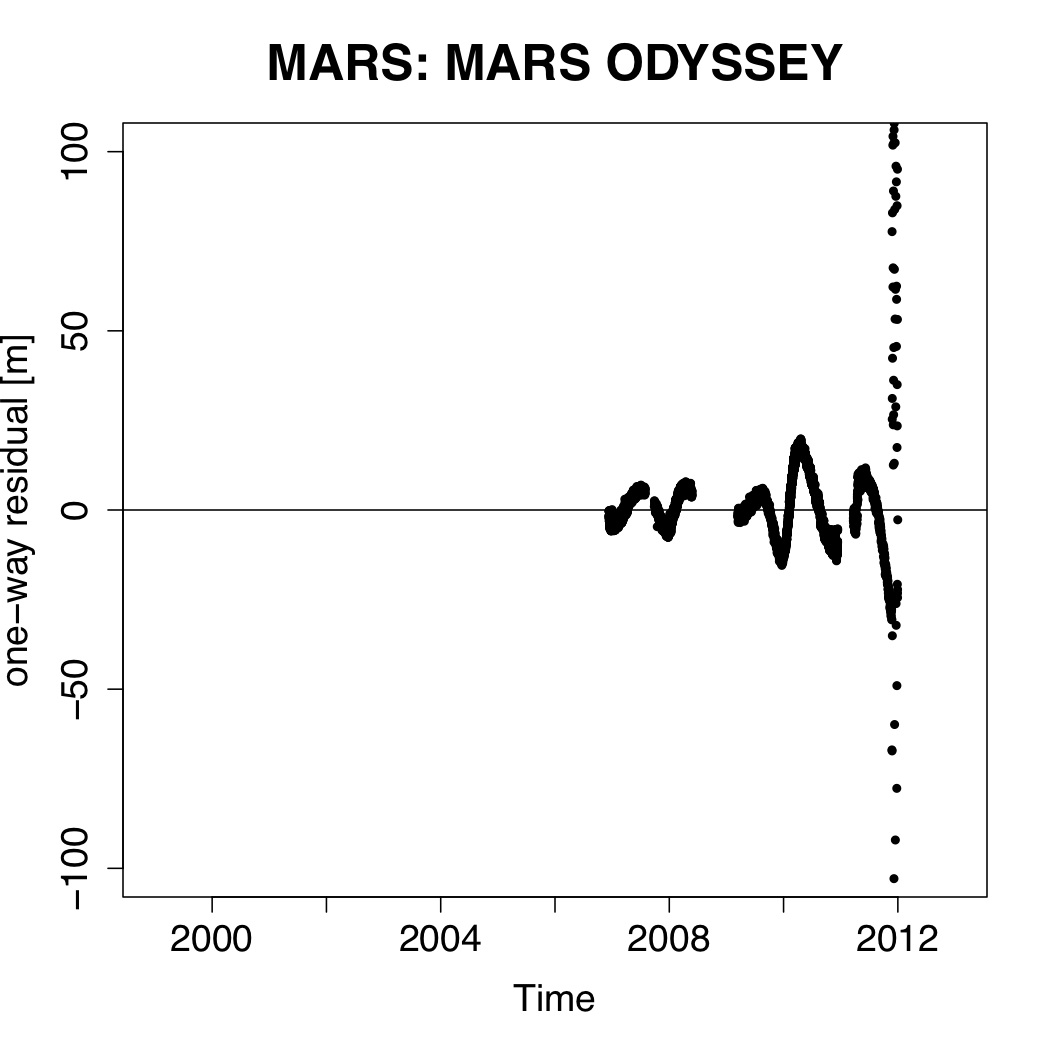}\\
\end{tabular}
\caption{Residuals for Mars Viking, Pathfinder, MGS and Mars Odyssey one-way range in meters obtained with INPOP10e, INPOP13c.}
\label{comparomc1}
\end{center}
\end{table}
%\end{figure}

\begin{table}
\vspace{-0.5cm}
\begin{center}
\begin{tabular}{c c}
%\hline
INPOP13c & INPOP10e \\
\includegraphics[scale=0.4]{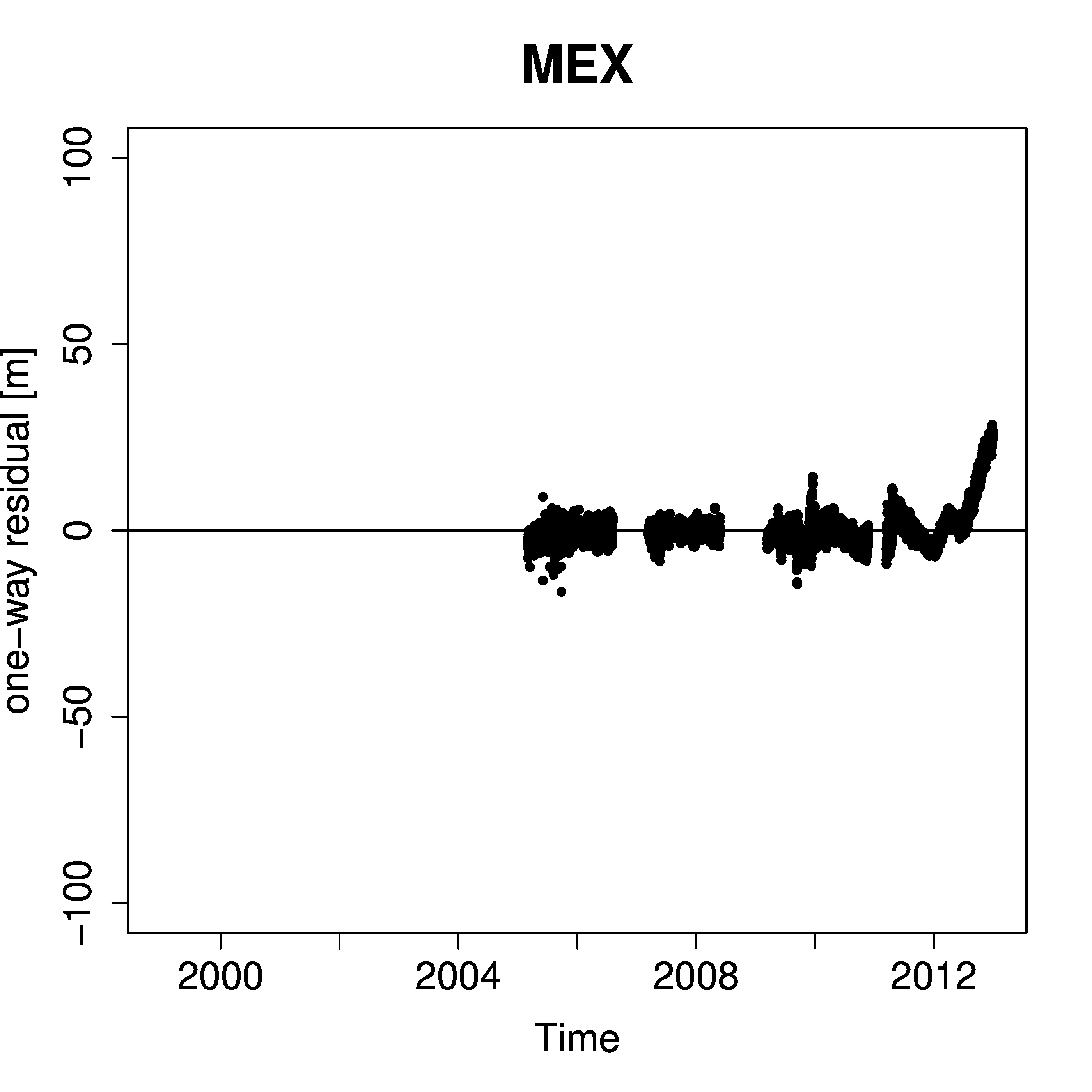}&\includegraphics[scale=0.4]{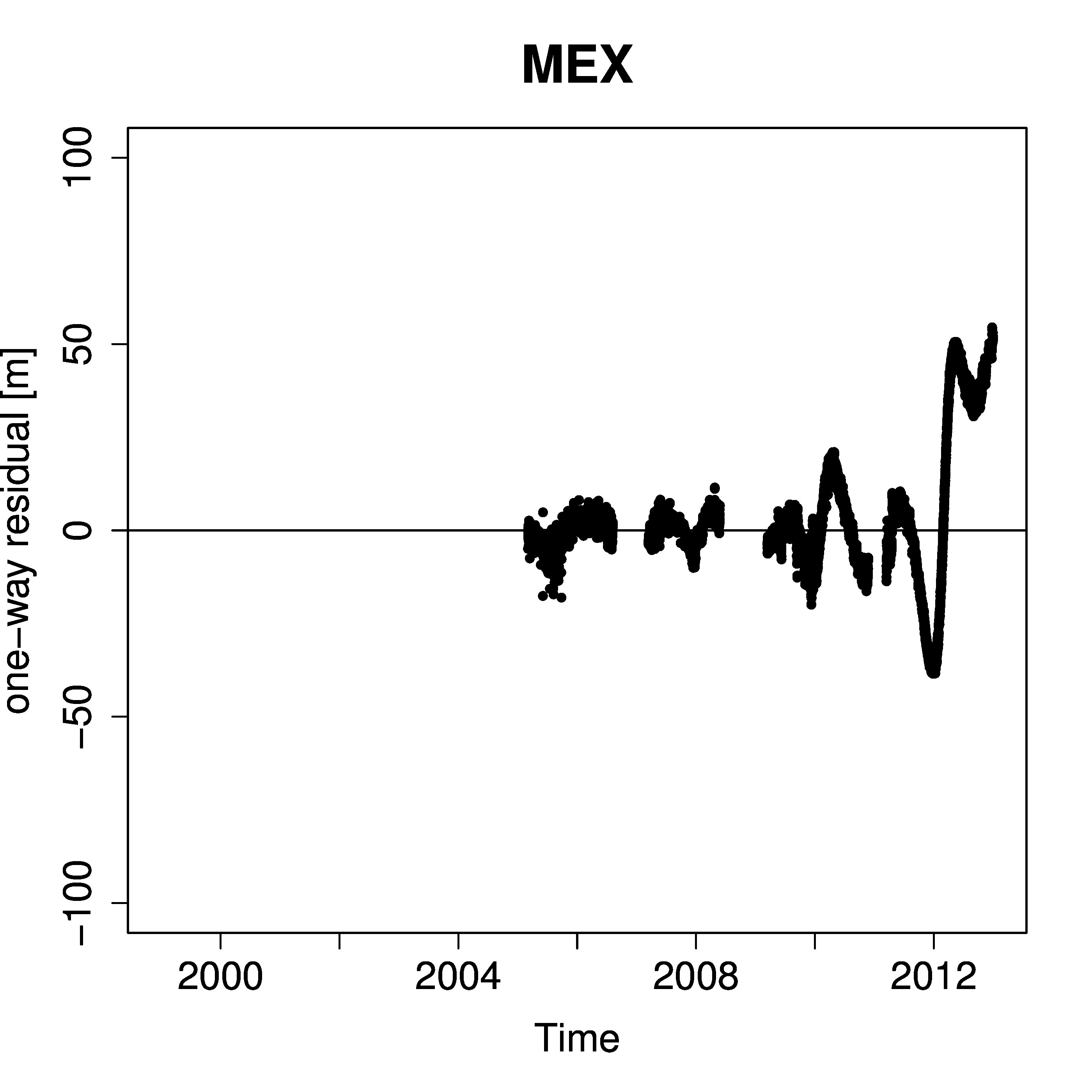}\\
\includegraphics[scale=0.4]{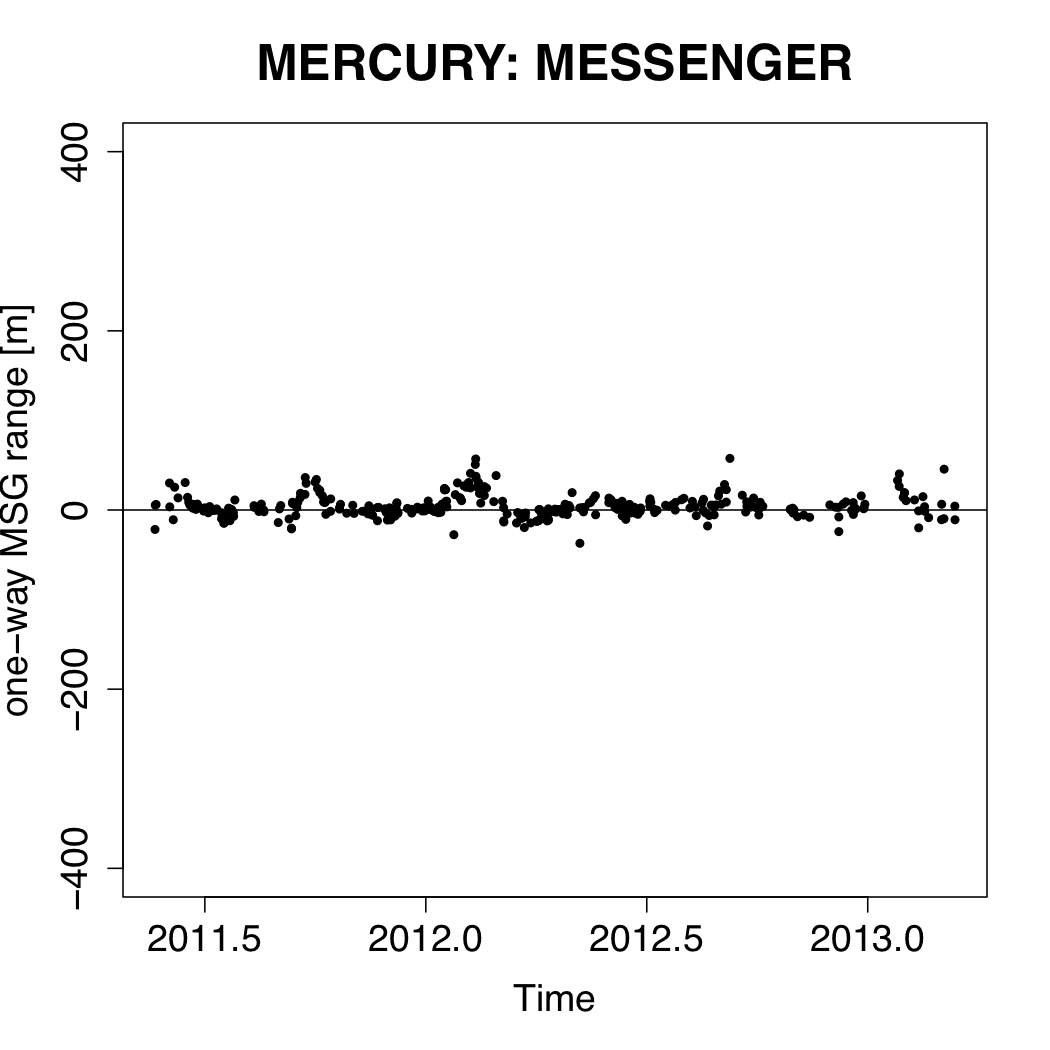}&\includegraphics[scale=0.4]{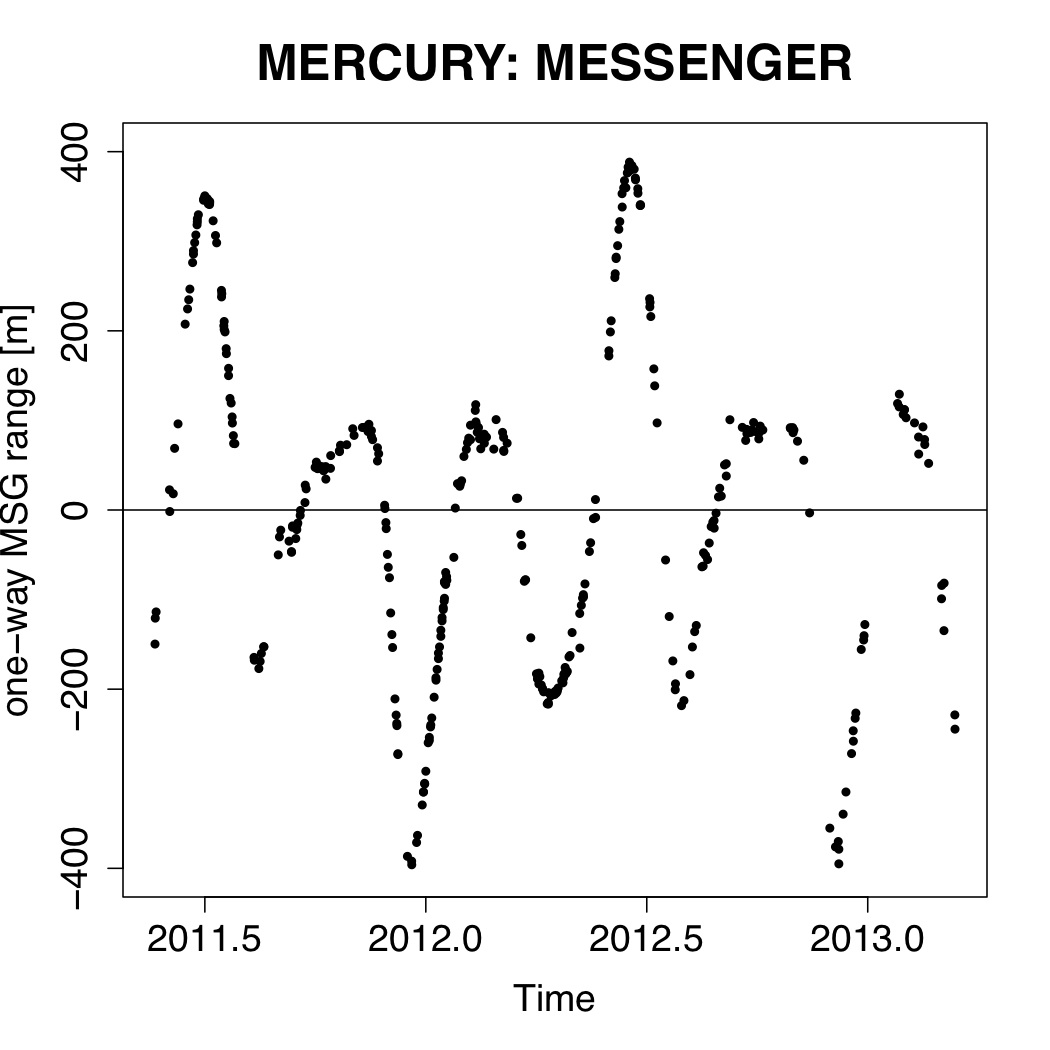}\\
\includegraphics[scale=0.4]{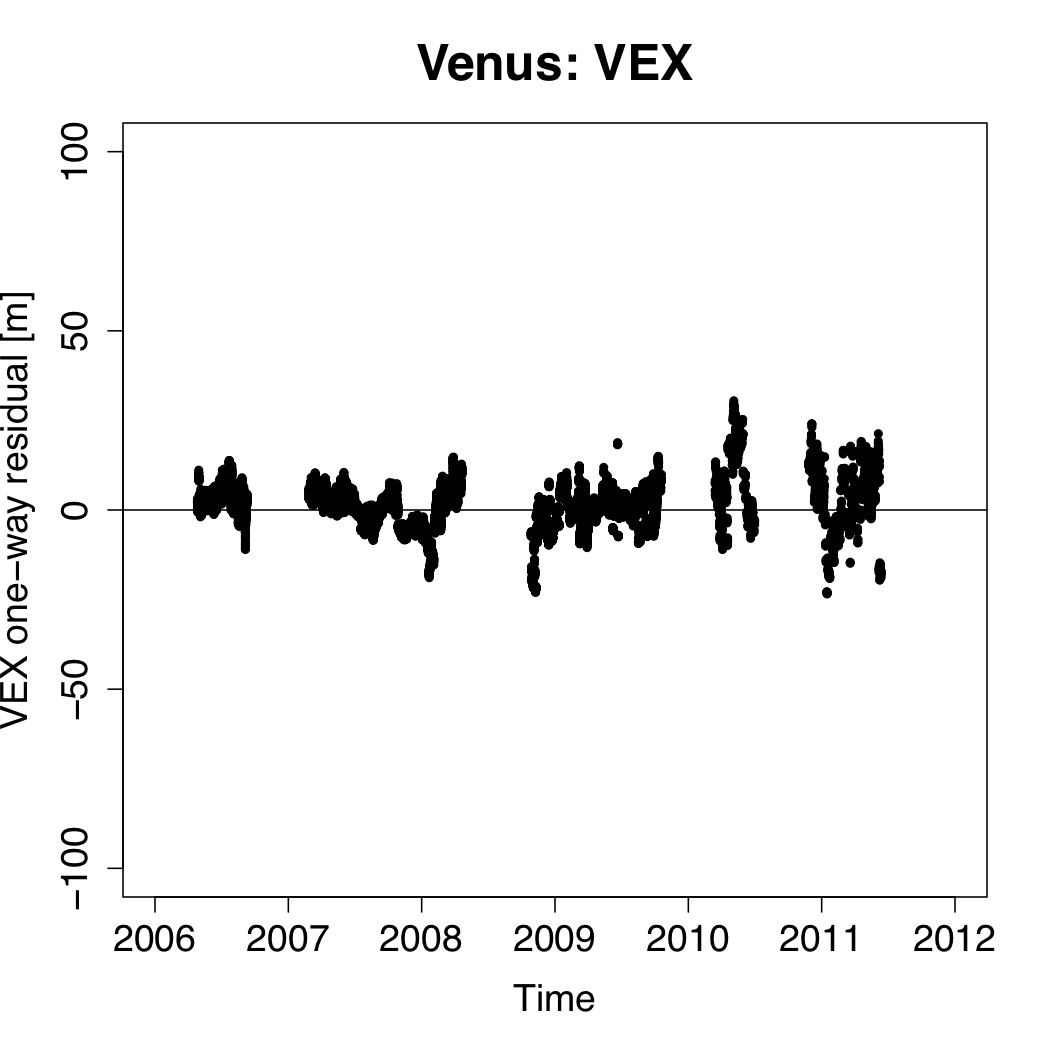}&\includegraphics[scale=0.4]{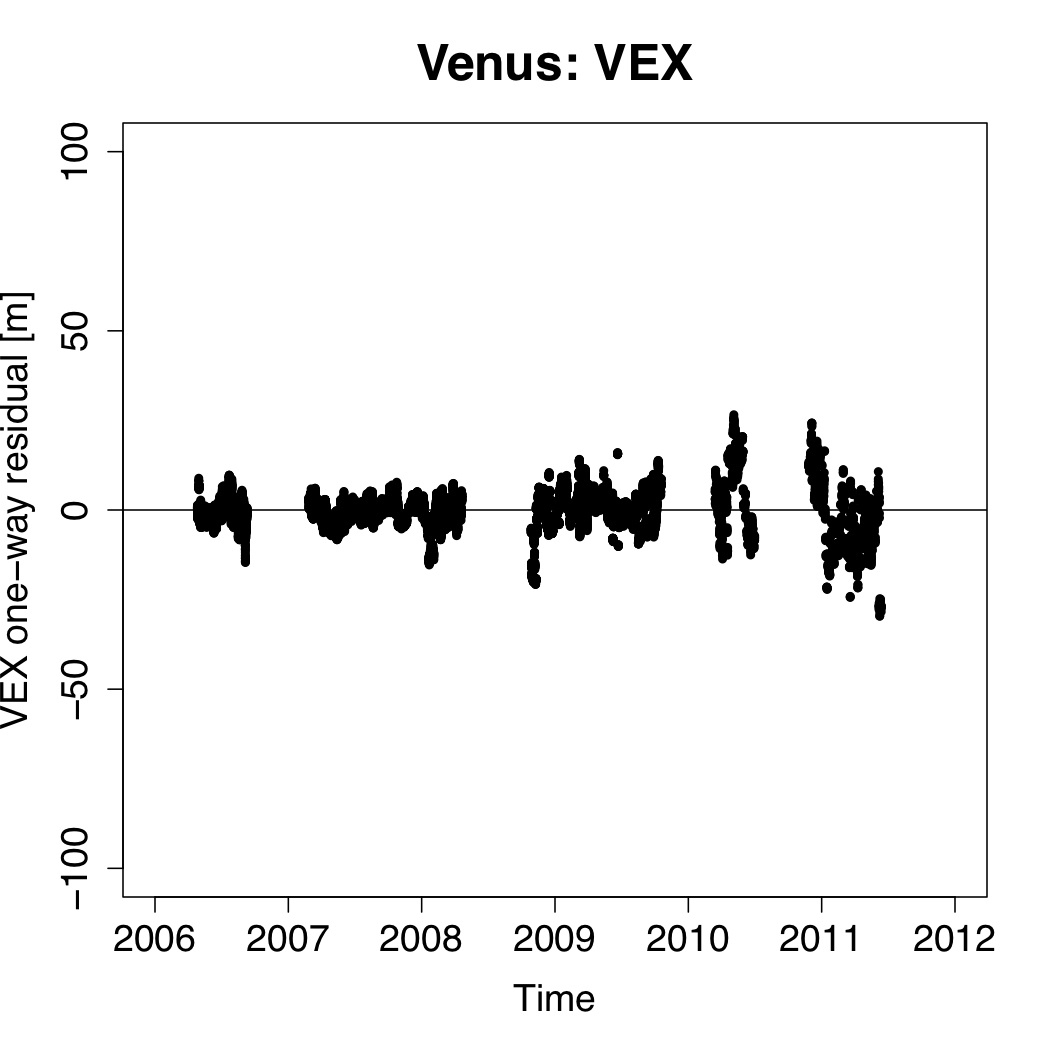}\\
\end{tabular}
\caption{Residuals for Mars MEX orbiter, Mercury MESSENGER and Venus VEX one-way range in meters obtained with INPOP13c and INPOP10e}
\label{comparomc2}
\end{center}
%\end{table}
\end{table}

\section{Asteroid masses}
\label{sect_ast}
\begin{figure}
\begin{center}
\includegraphics[scale=0.8]{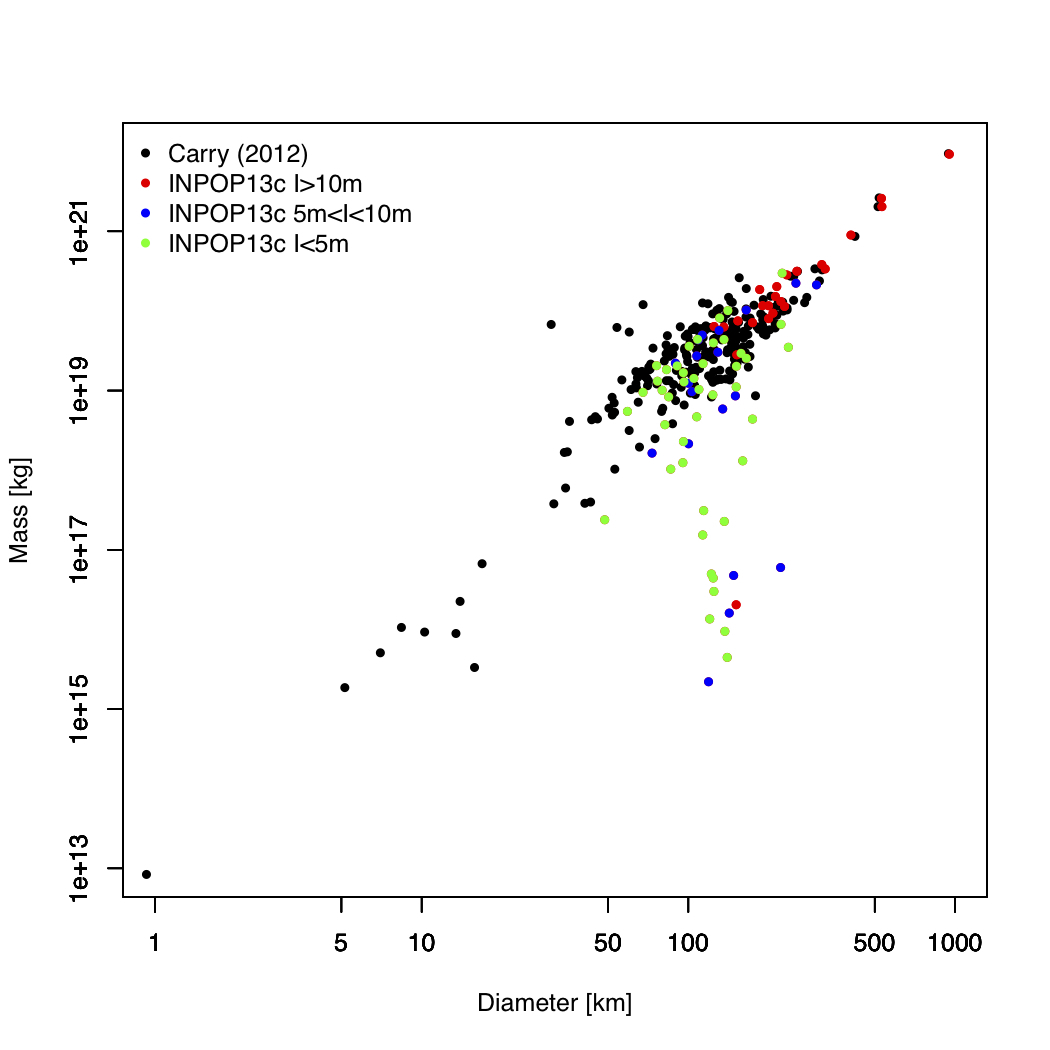}
\caption{INPOP13c Asteroid mass determination. The x-axis is the log of diameters in kilometers (as given in \cite{2010A&A...514A..96K}) when the y-axis is the log of the INPOP13c estimated masses in colors (see text) and of masses extracted from \cite{CARRY2012} in black. $I$ stands for the impact in meters of the asteroids on the Earth-Mars distances over the interval from 1970 to 2010 (see \cite{2010A&A...514A..96K}).}
\label{ast}
\end{center}
\end{figure}

139 asteroids are used in INPOP13c with 87 asteroid masses actually estimated during the fit, 52 being fixed to masses derived from taxonomic classes and diameter estimations or to better mass estimations deduced from binary system or spacecraft flybys. No asteroid ring is considered in INPOP13c.
Comparisons between the 87 asteroid masses obtained during the INPOP13c fit and masses collected by \cite{CARRY2012} are presented in Tables 10 and 11 and in Figure \ref{ast}.
Figure \ref{ast} pictures a log-log relation between asteroid masses and asteroid diameters as shown in \cite{CARRY2012} and \cite{2013Icar..226..723D}.
 Such a trend is a good tool for analyzing the quality of asteroid masses deduced from INPOP13c.
 One can first notice than asteroids having their masses deduced from planetary ephemerides have their diameters bigger than 50 kilometers. Only big objects can induce perturbations detectable thought inner planet observations.

 Three families of asteroids are furthermore be considered on Figure \ref{ast}: the big perturbers (in red) of the Earth-Mars distances inducing more than 10 meters on the geocentric distances of Mars over a 40-year interval (as defined in \cite{2010A&A...514A..96K}), the asteroids inducing a modification between 5 to 10 meters of the geocentric distances of Mars over the same interval of time (in blue) and the small perturbers (in green) modifying the Earth-Mars distances by less than 5 meters. As it appears clearly on Figure \ref{ast}, the estimations of the first two classes are quite compatible with the one obtained by \cite{CARRY2012} and follow generally well the trend with the diameters. The estimations of the smallest perturbers are, however, less compatible with the trend.

The 42 masses of the first two classes of asteroids are gathered in Table 10. They represent 48\% of the whole sample of the estimated masses and can be used for computing mean values of C, S, B and M densities as defined in \cite{2013Icar..226..723D} and mean cumulated masses for each of these classes as the results given in Table \ref{statdens}. The comparison to \cite{CARRY2012} obtained for the same sample of 42 masses confirms the consistency of the two samples. The comparison to estimations published by \cite{2013Icar..226..723D} and obtained for a whole population of asteroids indicates that most of the perturbing objects belong to the C and S taxonomic classes when a lack of M-class objects seems to correspond to less perturbing objects (small or far from Mars).

\section{Lunar Laser Ranging}
INPOP13c is fitted to LLR observations from 1969 to 2013.
Compared to INPOP10e, more than 700 observations from CERGA, 800 from Apollo and 50 from Matera have beed added in the fit.
The number of parameters fitted is now 66 instead of 65 for INPOP10e:
\begin{itemize}
 \item an offset for Apollo observations between december 2010 and april 2012 has been added
 \item because of its better significance (ratio between the fitted value and the formal error), the Moon's Love number $h_2$ is now fitted instead of Moon's coefficient of potential $C_{30}$ (fixed to LP150Q value)
\end{itemize}
The values of all fitted parameters are given in Tables \ref{Tab_valeurs_parametres_LLR_dyn_I13b}, \ref{Tab_valeurs_parametres_LLR_I13b_reflecteurs} and \ref{Tab_valeurs_parametres_LLR_I13b_stations}.
Their formal errors (1$\sigma$) come from the covariance matrix of the least square fit and can be much smaller than the physical uncertainties.

\begin{figure}
\begin{center}
\includegraphics[scale=0.8]{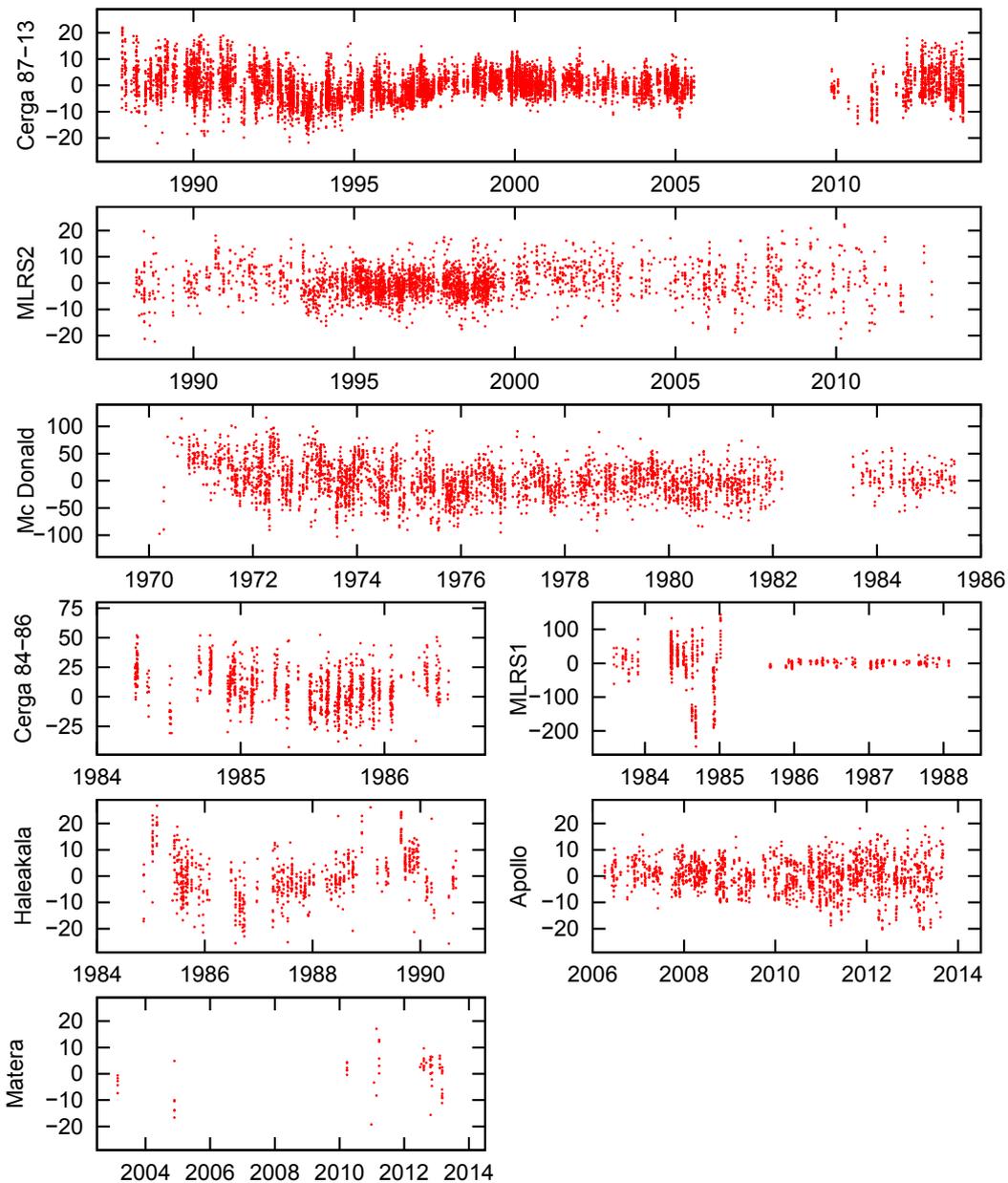}
\caption{Postfit LLR residuals with INPOP13c for each station, expressed in centimeters.}
\label{FigResidusLLR13b}
\end{center}
\end{figure}

On table \ref{Tab_residus_LLR_I13b} and figure \ref{FigResidusLLR13b} are given the LLR residuals.
The degradation on recent data was already noticed with INPOP10e, but it is amplified with INPOP13c.
For Cerga, the residuals grow from 4 to 6 centimeters between before and after 2009.
For Apollo, the standard deviation grows continuously from 5 to 7.6 centimeters between 2006 and 2013.
These increases of residuals would vanish with the fit of the tectonic plate motion for the corresponding stations, but we think the problem comes from the dynamical model, with a lack that can be partly compensated by a change in station motion.

On table \ref{Tab_comp_INPOP_JPL} is shown a comparison of residuals between INPOP13c and several JPL's solutions. JPL's residuals are here obtained by applying the same reduction process (model of light propagation) to the planetary and lunar motions of the corresponding JPL's solution. Only parameters involved in the reduction (see tables \ref{Tab_valeurs_parametres_LLR_I13b_reflecteurs} and \ref{Tab_valeurs_parametres_LLR_I13b_stations}) are refitted. Residuals computed here are then certainly different from the ones computed directly by the JPL.
One can see for DE421 a small degradation on Apollo data (from 4.5 cm to 5.1 cm); this solution was released in 2008 and it is normal to observe such a degradation on data that were not available at this time, and thus not used to constrain the parameters.
It should be stressed that DE421 residuals are better than the ones of INPOP13c (even on recent data); it thus demonstrates the better quality (extrapolation capabilities) of DE421 dynamical model.
For DE430, residuals are improved relativ to the ones of DE421; the dynamical model is slightly different (Earth tides, degrees of Moon's and Earth's potential) and  it is fitted to the latest data.
The bad behaviour of INPOP13c compared to JPL's solutions is thus certainly due to a lack in the dynamical model (missing interaction?).
The main difference between INPOP and JPL's solutions is the presence (since DE418) of a lunar liquid core interacting with the mantle.
Its equations are given in \cite[eqs. 6,7,8]{Williams2008}:
\begin{equation}\label{Eq_noyau_Lune}
\left \{
\begin{array}{l}
\frac {d}{dt} I_m \vec{\omega}_m + \vec{\omega}_m \wedge I_m \vec{\omega}_m = \vec{T}_g + \vec{T}_{cmb} \\
\frac {d}{dt} I_f \vec{\omega}_f + \vec{\omega}_m \wedge I_f \vec{\omega}_f = - \vec{T}_{cmb} \\
\vec{T}_{cmb} = K_{\nu}\left( \vec{\omega}_f - \vec{\omega}_m \right)  
               + \left( C_f-A_f\right) \left( \tilde z \cdot \vec{\omega}_f\right)\left(  \tilde z \wedge \vec{\omega}_f  \right)\\
\end{array}
\right .
\end{equation}
In these expressions,
\begin{itemize}
 \item $I_m$ and $I_f$ are respectively the inertia matrices of the mantle and the fluid core
 \item $\vec{\omega}_m$ and $\vec{\omega}_f$ are the instant rotation vectors of the mantle and the fluid core
 \item $\vec{T}_g$ is the gravitational torque exerted by external bodies (Sun, Earth, ...)
 \item $\vec{T}_{cmb}$ is the torque due to interactions at core/mantle boundary
 \item $A_f$ and $C_f$ are the moments of inertia of the axisymetrical fluid core
 \item $\tilde z$ is the direction of the pole of the mantle
 \item $K_{\nu}$ is a friction coefficient at the core/mantle boundary
\end{itemize}

Analysis of DE431 residuals are then interesting.
This solution, similar to DE430, is briefly described in \cite{2014IPNPR.196C...1F} and \cite{2014IPNPR.196C...1F}.
In order to provide a long term solution (especially in the past), the friction torque at the lunar core/mantle boundary has been neglected ($ K_{\nu}=0 $ in eq. \ref{Eq_noyau_Lune}). In table \ref{Tab_comp_INPOP_JPL}, one can see that its behaviour on recent data is the same as INPOP13c's one, with standard deviations of residuals increasing with time for both Apollo and Cerga; the values reached are often greater than with INPOP13c.
This indicates that not only a lunar core is necessary, but also that the friction torque at the core/mantle boudary must not be neglicted.

INPOP13c does not take into account any lunar core yet, even if expressions \ref{Eq_noyau_Lune} are already implemented in the equations of motions.
The problem is that with fixed values of physical parameters ($I_f$, $C_f-A_f$ and $K_{\nu}$) close to DE421, DE423 or DE430 ones, after fit of 69 other parameters (the 66 ones fitted in INPOP13c and the initial conditions of the core's instant vector of rotation), there is no significant improvement on residuals. This is why the Lunar core has not been activated for INPOP13c.

In conclusion, INPOP13c is not a solution as good as DE430 (or DE421) on LLR observations, even if it is improved compared to INPOP10e (fitted to more recent data).
A work is still in progress in order to determine optimal values for the physical parameters of the Lunar core, certainly farther away from JPL's ones.

 \begin{table}
 \caption{Cumulated masses of the 42 most perturbing objects and deduced mean densities for taxonomic classes defined in \cite{2013Icar..226..723D}.}
 \label{statdens}
 \begin{tabular}{c c c c c c}
 \hline
 & C-Class & S-Class & B-Class & M-Class & V-Class \\
 \hline
 This paper & & & & \\
 Cumulated mass [kg]& 1.1973e+21 & 1.61e+20 &  2.04e+20& 2.2e+19 & 2.590e+20\\
 \% of total mass & 65\% & 9 \% &  11 \% & 1.2 \% & 14 \%\\
 \% without CPVB & 14 \% & 9 \% & 0 \% & 1.2\% & 0 \% \\
 Density  [g.cm$^{-3}$]& 2.19 $\pm$ 1.89 & 2.89 $\pm$ 1.79 & 2.59 & 2.40 $\pm$ 2.26 &  3.32\\
 \hline
 (DeMeo and Carry 2013)\cite{2013Icar..226..723D} & & & & \\
 Cumulated mass [kg]& 1.42e+21 & 2.27e+20 & 3.0e+20 &  8.82+19 & 2.59e+20\\
 \% of total mass & 52\% & 8.4 \% & 11.10 \% &3.26 \% & 9.59 \%\\
  \% without CPVB & 14 \% & 8.4 \% & 3.5 \% & 3.26 \%& 0.01 \%\\
 Density  [g.cm$^{-3}$]& 1.33 $\pm$ 0.58 & 2.72 $\pm$ 0.54 & 2.38 $\pm$ 0.45 & 3.49 $\pm$ 1.00 & 1.93 $\pm$ 1.07\\
 \hline
 (Carry 2012)\cite{CARRY2012} & & & \\
  Cumulated mass [kg]& 1.18864e+21 & 1.5007e+20 & 2.04e+20 &  2.863e+19& 2.63e+20\\
 \% of total mass & 65\% & 8.2 \% & 11.10 \% & 1.56 \% & 14 \%\\
  \% without CPVB & 13\% & 8.2\% & 0 \% & 1.56\%& 0 \%\\
 Density  [g.cm$^{-3}$]& 1.54 $\pm$ 0.55 & 2.85 $\pm$ 1.04 & 2.30 & 4.53 $\pm$ 2.38  & 3.00\\
 \hline
 \end{tabular}
 \end{table}

%\begin{figure}
%\begin{center}
%\includegraphics[scale=1.0]{/Users/Agnes/INPOP/comparfort987/de423.jpeg}
%\caption{Postfit and extrapolated residuals obtained with DE423}
%\end{center}
%\end{figure}

%\label{secapp}
%%---------------------------------

%%---------------------

%\subsection{Asteroid masses}

%Hyperlinks can be introduced as follows: \url{http://www.sf2a.asso.fr/}.

\begin{landscape}

\begin{table}
\caption{Statistics of the residuals obtained after the INPOP13c fit for common data sample between INPOP13c and INPOP10e. For comparison, means and standard deviations of residuals obtained with INPOP10e. }
\begin{center}
\begin{tabular}{l c l c | c c | c c }
\hline
Type of data & & Nbr & Time Interval & \multicolumn{2}{c}{INPOP10e}& \multicolumn{2}{c}{INPOP13c} \\
\hline
Mercury & range [m]& 462 & 1971.29 - 1997.60 & -45.3 & 872.499 &  -101.524 &    861.494   \\
Mercury  Mariner & range [m]& 2 & 1974.24 - 1976.21 & -31.850 &  109.191 & -196.405 &     19.636   \\
Mercury  flybys  Mess & ra [mas]& 3 & 2008.03 - 2009.74 & 0.738  & 1.485 & 0.901 &      1.355   \\
Mercury  flybys  Mess & de [mas]& 3 & 2008.03 - 2009.74 & 2.422  & 2.517 & 2.472 &       2.408  \\
Mercury  flybys  Mess & range [m]& 3 & 2008.03 - 2009.74 & -5.067 & 5.804  & 3.190 &       7.699   \\
\hline
Venus & VLBI [mas]& 46 & 1990.70 - 2010.86 & 1.590 & 2.602  & 1.591 &      2.575    \\
Venus & range [m]& 489 & 1965.96 - 1990.07 & 500.195 & 2234.924  & 504.569 &   2237.636   \\
Venus  Vex & range [m]& 22145 & 2006.32 - 2009.78 & -0.004 & 4.093  &  1.042 &      5.079  \\
\hline
Mars & VLBI [mas]& 96 & 1989.13 - 2007.97 & -0.004 & 0.407  &  0.018 &      0.418  \\
Mars  Mex & range [m]& 13842 & 2005.17 - 2009.78 & 0.485 & 3.189 &  -0.465 &      1.840  \\
Mars  MGS & range [m]& 13091 & 1999.31 - 2006.83 & -0.341 & 3.805  & 0.362 &      3.777   \\
Mars  Ody & range [m]& 5664 & 2006.95 - 2010.00 & 0.280 & 4.155  &   1.55 &      2.30  \\
Mars  Path & range [m]& 90 & 1997.51 - 1997.73 & -6.289 & 13.663  & 19.324 &     14.096   \\
Mars  Vkg & range [m]& 1257 & 1976.55 - 1982.87 & -1.391 & 39.724  & -1.494 &     41.189   \\
\hline
Jupiter & VLBI [mas]& 24 & 1996.54 - 1997.94 & -0.291 & 11.068  & -0.450 &     11.069    \\
Jupiter & ra [arcsec]& 6532 & 1914.54 - 2008.49 & -0.039 & 0.297 &  -0.039 &      0.297   \\
Jupiter & de [arcsec]& 6394 & 1914.54 - 2008.49 & -0.048 & 0.301  & -0.048 & 0.301   \\
Jupiter  flybys & ra [mas]& 5 & 1974.92 - 2001.00 & 2.368 & 3.171  & 2.554 &      2.961   \\
Jupiter  flybys & de [mas]& 5 & 1974.92 - 2001.00 & -10.825 & 11.497 & -10.853 &      11.425  \\
Jupiter  flybys & range [m]& 5 & 1974.92 - 2001.00 & -907.0 & 1646.210 &  -985.957 &    1775.627  \\
\hline
Saturne & ra [arcsec]& 7971 & 1913.87 - 2008.34 & -0.006 & 0.293 &  -0.006 & 0.293  \\
Saturne & de [arcsec]& 7945 & 1913.87 - 2008.34 & -0.012 & 0.266 & -0.012 & 0.266  \\
Saturne  VLBI  Cass & ra [mas]& 10 & 2004.69 - 2009.31 & 0.215 & 0.637 &  0.113 &      0.630   \\
Saturne  VLBI  Cass & de [mas]& 10 & 2004.69 - 2009.31 & 0.280 & 0.331 &   -0.115 &      0.331 \\
Saturne  Cassini & ra [mas]& 31 & 2004.50 - 2007.00 & 0.790 & 3.879 &  0.663 &      3.883  \\
Saturne  Cassini & de [mas]& 31 & 2004.50 - 2007.00 & 6.472 & 7.258 &   5.906 &       7.284 \\
Saturne  Cassini & range [m]& 31 & 2004.50 - 2007.00 & -0.013 & 18.844 &   0.082 &      23.763   \\
\hline
\end{tabular}
\end{center}
\label{omctab0}
\end{table}

\begin{table}
\caption{Statistics of the residuals obtained after the INPOP13c fit for common data sample between INPOP13c and INPOP10e. For comparison, means and standard deviations of residuals obtained with INPOP10e are given. }
\begin{center}
\begin{tabular}{l c l c | c c | c c }
\hline
Type of data & & Nbr & Time Interval & \multicolumn{2}{c}{INPOP10e}& \multicolumn{2}{c}{INPOP13c} \\
\hline
Uranus & ra [arcsec]& 13016 & 1914.52 - 2011.74 & 0.007 & 0.205 & 0.007 & 0.205  \\
Uranus & de [arcsec]& 13008 & 1914.52 - 2011.74 & -0.006 & 0.234 & -0.006 & 0.234   \\
Uranus  flybys & ra [arcsec]& 1 & 1986.07 - 1986.07 & -0.021 & 0.000 & -0.021 & 0.000  \\
Uranus  flybys & de [arcsec]& 1 & 1986.07 - 1986.07 & -0.028 & 0.000 & -0.028 & 0.000    \\
Uranus  flybys & range [m]& 1 & 1986.07 - 1986.07 & 19.738 & 0.000 &20.771 & 0.000  \\
\hline
Neptune & ra [arcsec]& 5395 & 1913.99 - 2007.88 & 0.000 & 0.258 & 0.003 & 0.258 \\
Neptune & de [arcsec]& 5375 & 1913.99 - 2007.88 & -0.000 & 0.299 & -0.002 & 0.299  \\
Neptune  flybys & ra [arcsec]& 1 & 1989.65 - 1989.65 & -0.012 & 0.000 & -0.011& 0.000    \\
Neptune  flybys & de [arcsec]& 1 & 1989.65 - 1989.65 & -0.005 & 0.000 & -0.005 & 0.000    \\
Neptune  flybys & range [m]& 1 & 1989.65 - 1989.65 & 69.582 & 0.000 & 51.507& 0.000     \\
\hline
Pluto & ra [arcsec]& 2458 & 1914.06 - 2008.49 & 0.034 & 0.654 & 0.020 & 0.574     \\
Pluto & de [arcsec]& 2462 & 1914.06 - 2008.49 & 0.007 & 0.539 & 0.001 & 0.525     \\
Pluto  Occ & ra [arcsec]& 13 & 2005.44 - 2009.64 & 0.003 & 0.047 & -0.100 &      0.044     \\
Pluto  Occ & de [arcsec]& 13 & 2005.44 - 2009.64 & -0.006 & 0.018 &  0.000 &      0.027  \\
Pluto  HST & ra [arcsec]& 5 & 1998.19 - 1998.20 & -0.033 & 0.043 & -0.018 & 0.044     \\
Pluto  HST & de [arcsec]& 5 & 1998.19 - 1998.20 & 0.028 & 0.048 & -0.026 & 0.048   \\
\hline
\hline
\end{tabular}
\end{center}
\label{omctab1}
\end{table}

\begin{table}
\caption{Statistics of INPOP13c  postfit residuals for new samples included in the fit. For comparison, means and standard deviations of residuals obtained with INPOP10e on these prolongated intervals are also given.}
\begin{center}
\begin{tabular}{l c l c | c c | c c }
\hline
Type of data & & Nbr & Time Interval & \multicolumn{2}{c}{INPOP10e}& \multicolumn{2}{c}{INPOP13c} \\
\hline
Mercure Messenger & range [m]& 371 &  2011.39 - 2013.20 &  7.239   & 189.739   &  4.008 &     12.387\\
\hline
Venus Vex & range [m]& 2825 &  2009.78 - 2011.45 & 1.84 &  16.54 &    5.12 &     15.70 \\
\hline
Mars Mex & range [m]& 12268 &  2009.78 - 2013.00 &   3.646  &  23.296   & 1.234 &     5.571 \\
Mars Ody & range [m]& 3510 &  2010.00 - 2012.00 &    4.834   &  9.586    & 0.741&     1.778 \\
\hline
\end{tabular}
\end{center}
\label{omcsupp}
\end{table}

\end{landscape}

\begin{landscape}
\begin{table}
\caption{Asteroid masses obtained with INPOP13c and compared with values extracted from \cite{CARRY2012}. This table gathers the asteroids inducing a maximum change of more than 5 meters on the Earth-Mars distances on a [1970:2010] interval. Column 4 gives the value of this maximum impact of each asteroid on the Earth-Mars distances as defined in \cite{2010A&A...514A..96K}. The Column 5 gives values of diameters obtained from IRAS and WISE surveys as given by \cite{2010A&A...514A..96K} and Columns 6 and 7 present the masses and their uncertainties estimated by \cite{CARRY2012}. The last class gives the taxonomical index extracted from the Tholen classification (\cite{Tholen1984}).}
%\vspace{-0.5cm}
\begin{center}
\label{mass0}
\begin{tabular}{c c c c c c c c}
\hline
IAU designation & INPOP13c & 1-$\sigma$ & Impact & diam & Carry 2012 & 1-$\sigma$ & T-Class\\
& kg & kg & m & km & kg & kg & \\
4 & 2.590564e+20 & 1.157098e+18 & 1198.9530440 & 530.00 & 2.63e+20 & 5.00e+18 & V\\
1 & 9.291844e+20 & 3.844558e+18 & 793.7412482 & 952.40 & 9.44e+20 & 6.00e+18 & G\\
2 & 2.041335e+20 & 2.714066e+18 & 146.2695538 & 532.00 & 2.04e+20 & 4.00e+18 & B\\
324 & 1.138045e+19 & 5.419632e+17 & 93.5357526 & 229.44 & 1.03e+19 & 1.00e+18 & CP\\
10 & 8.980667e+19 & 7.853611e+18 & 77.0025537 & 407.12 & 8.63e+19 & 5.20e+18 & C\\
19 & 8.003182e+18 & 9.346277e+17 & 59.0689816 & 200.00 & 8.60e+18 & 1.46e+18 & G\\
3 & 2.834174e+19 & 1.300802e+18 & 55.6392528 & 233.92 & 2.73e+19 & 2.90e+18 & S\\
704 & 3.821399e+19 & 4.131905e+18 & 34.4922565 & 316.62 & 3.28e+19 & 4.50e+18 & F\\
532 & 1.310927e+19 & 1.512167e+18 & 32.7144140 & 222.39 & 1.15e+19 & 2.80e+18 & S\\
9 & 1.165527e+19 & 1.205371e+18 & 29.6062727 & 190.00 & 8.39e+18 & 1.67e+18 & S\\
7 & 1.162482e+19 & 9.666165e+17 & 27.8218292 & 199.83 & 1.29e+19 & 2.10e+18 & S\\
29 & 1.517009e+19 & 1.903232e+18 & 26.6731763 & 212.22 & 1.29e+19 & 2.00e+18 & S\\
13 & 9.372223e+18 & 2.364485e+18 & 22.0382227 & 207.64 & 8.82e+18 & 4.25e+18 & G\\
15 & 3.149656e+19 & 1.569941e+18 & 21.5545880 & 255.33 & 3.14e+19 & 1.80e+18 & S\\
6 & 1.855289e+19 & 1.282286e+18 & 21.1504947 & 185.18 & 1.39e+19 & 1.00e+18 & S\\
14 & 2.813508e+18 & 9.790674e+17 & 18.1615646 & 152.00 & 2.91e+18 & 1.88e+18 & S\\
11 & 7.499181e+18 & 1.387315e+18 & 17.3007697 & 153.33 & 5.91e+18 & 4.50e+17 & S\\
8 & 6.288016e+18 & 7.000950e+17 & 12.6635308 & 135.89 & 9.17e+18 & 1.75e+18 & S\\
45 & 2.021796e+19 & 2.393618e+18 & 11.7902641 & 214.63 & 5.79e+18 & 1.40e+17 & FC\\
41 & 7.131347e+18 & 6.700157e+17 & 11.5682732 & 174.00 & 6.31e+18 & 1.10e+17 & C\\
405 & 6.376346e+18 & 3.743249e+17 & 11.3784687 & 124.90 & 1.38e+18 & 1.40e+17 & C\\
145 & 2.042610e+15 & 2.192145e+15 & 11.0486157 & 151.14 & 2.08e+18 & 5.70e+17 & C\\
511 & 3.371614e+19 & 5.716552e+18 & 10.2480612 & 326.06 & 3.38e+19 & 1.02e+19 & C\\
52 & 2.124138e+19 & 4.905582e+18 & 9.8407320 & 302.50 & 2.38e+19 & 5.80e+18 & CF\\
16 & 2.229501e+19 & 3.642799e+18 & 9.7005515 & 253.16 & 2.72e+19 & 7.50e+18 & M\\
419 & 3.030678e+18 & 9.012887e+17 & 9.5851372 & 129.01 & 1.72e+18 & 3.40e+17 & F\\
23 & 2.730810e+18 & 5.714392e+17 & 9.0667512 & 107.53 & 1.96e+18 & 9.00e+16 & S\\
488 & 8.576408e+17 & 3.383274e+18 & 8.6136830 & 150.13 & 2.48e+18 & 1.14e+18 & C\\
230 & 4.342298e+18 & 1.927946e+18 & 7.6195204 & 108.99 & 1.89e+18 & 1.90e+17 & S\\
\hline
 \end{tabular}
\end{center}

\end{table}

\begin{table}
%\begin{center}
\begin{center}
\begin{tabular}{c c c c c c c c}
\hline
IAU designation & INPOP13c & 1-$\sigma$ & Impact & diam & Carry 2012 & 1-$\sigma$ & T-Class\\
& kg & kg & m & km & kg & kg & \\
187 & 5.686063e+18 & 7.819741e+17 & 7.5924498 & 130.40 & 1.80e+18 & 8.50e+17 & C\\
111 & 5.869674e+17 & 1.980823e+18 & 6.9848015 & 134.55 & 1.76e+18 & 4.40e+17 & C\\
109 & 2.209681e+18 & 5.165756e+17 & 6.8652431 & 89.44 & NA & NA & GC\\
42 & 2.145282e+17 & 6.694597e+17 & 6.8289220 & 100.20 & 1.58e+18 & 5.20e+17 & S\\
63 & 9.693634e+17 & 2.797805e+17 & 6.4510297 & 103.14 & 1.53e+18 & 1.50e+17 & S\\
12 & 4.976460e+18 & 4.138827e+17 & 6.1587944 & 112.77 & 2.45e+18 & 4.60e+17 & S\\
144 & 1.605988e+15 & 1.992285e+15 & 6.0865220 & 142.38 & 5.30e+18 & 1.20e+18 & C\\
5 & 2.204297e+14 & 4.293857e+17 & 5.5329555 & 119.07 & 2.64e+18 & 4.40e+17 & S\\
48 & 5.994443e+15 & 3.760429e+16 & 5.3602865 & 221.80 & 6.12e+18 & 2.96e+18 & CG\\
59 & 1.025852e+19 & 3.278702e+16 & 5.3247597 & 164.80 & 3.00e+18 & 5.00e+17 & CP\\
30 & 1.233848e+18 & 2.117636e+18 & 5.3223063 & 100.15 & 1.74e+18 & 4.90e+17 & S\\
51 & 4.764145e+15 & 5.488586e+17 & 5.1088310 & 147.86 & 2.48e+18 & 8.60e+17 & CU\\
516 & 1.638896e+17 & 7.806427e+17 & 5.0514023 & 73.10 & 1.43e+18 & 1.33e+18 & M\\
\hline
 \end{tabular}
\end{center}
\label{mass0}
\end{table}
 \end{landscape}

\begin{table}
\begin{center}
\caption{Asteroid masses obtained with INPOP13c and compared with values extracted from \cite{CARRY2012}. This table gathers more specifically the asteroids inducing a maximum change of less than 5 meters on the Earth-Mars distances on a [1970:2015] interval.}
%\vspace{-0.4cm}
\begin{tabular}{c c c c c c c c c}
\hline
IAU designation & INPOP13c & 1-$\sigma$ & Impact & diam & Carry 2013 & 1-$\sigma$\\
& kg & kg & m & km & kg & kg \\
89 & 2.017804e+18 & 1.158577e+18 & 4.8149336 & 151.46 & 6.71e+18 & 1.82e+18\\
451 & 2.979651e+19 & 6.217767e+18 & 4.7418532 & 224.96 & 1.09e+19 & 5.30e+18\\
313 & 1.283921e+18 & 7.546575e+17 & 4.7044991 & 96.34 & NA & NA\\
554 & 2.286580e+17 & 5.142521e+17 & 4.7036067 & 95.87 & 6.59e+17 & 6.60e+16\\
107 & 6.787732e+18 & 2.890148e+18 & 4.6303383 & 222.62 & 1.12e+19 & 3.00e+17\\
65 & 3.495611e+18 & 1.697773e+18 & 4.5364039 & 237.26 & 1.36e+19 & 3.10e+18\\
21 & 1.677741e+18 & 1.476184e+18 & 4.5269608 & 95.76 & 1.70e+18 & 1.00e+16\\
694 & 2.041762e+18 & 8.867518e+17 & 4.1874011 & 90.78 & NA & NA\\
120 & 4.385643e+17 & 5.288536e+18 & 3.9064935 & 174.10 & NA & NA\\
46 & 3.971305e+18 & 1.247291e+18 & 3.5594356 & 124.14 & 5.99e+18 & 5.00e+17\\
37 & 4.399527e+18 & 3.294685e+17 & 3.5514669 & 108.35 & NA & NA\\
216 & 4.422420e+15 & 6.267640e+17 & 3.4827870 & 124.00 & 4.64e+18 & 2.00e+17\\
164 & 1.420331e+18 & 2.912094e+17 & 3.4332830 & 104.87 & 9.29e+17 & 7.76e+17\\
410 & 8.832544e+17 & 2.119148e+17 & 3.3882431 & 123.57 & 6.24e+18 & 3.00e+17\\
56 & 1.536408e+16 & 5.755360e+17 & 3.2535822 & 113.24 & 4.61e+18 & 0.00e+00\\
34 & 2.178007e+18 & 3.787477e+17 & 3.2274053 & 113.54 & 3.66e+18 & 3.00e+16\\
95 & 4.382951e+18 & 6.566998e+17 & 3.0162129 & 136.04 & NA & NA\\
268 & 4.455482e+14 & 2.584680e+14 & 2.8481781 & 139.89 & 3.25e+18 & 2.26e+18\\
814 & 1.032312e+18 & 1.703245e+18 & 2.8300815 & 109.56 & NA & NA\\
209 & 1.310480e+17 & 6.530118e+18 & 2.5293436 & 159.94 & 4.59e+18 & 7.42e+18\\
337 & 5.473158e+17 & 1.595918e+17 & 2.4044223 & 59.11 & 1.08e+18 & 1.60e+17\\
386 & 2.542494e+18 & 1.501009e+18 & 2.3365190 & 165.01 & 8.14e+18 & 1.58e+18\\
739 & 4.680385e+17 & 1.611940e+18 & 2.2048585 & 107.53 & 1.16e+18 & 1.07e+18\\
602 & 3.002039e+15 & 2.484931e+15 & 2.0856729 & 124.72 & 1.02e+19 & 5.00e+17\\
141 & 8.167383e+18 & 1.558767e+18 & 1.9991831 & 131.03 & 8.25e+18 & 5.77e+18\\
804 & 2.932829e+18 & 9.832968e+17 & 1.9193459 & 157.58 & 5.00e+18 & 1.78e+18\\
776 & 1.110207e+18 & 8.848554e+17 & 1.8941086 & 151.17 & 2.20e+18 & 2.71e+18\\
127 & 4.973809e+15 & 1.581185e+16 & 1.7862193 & 122.00 & 3.08e+18 & 1.35e+18\\
308 & 1.010230e+19 & 2.356410e+18 & 1.7849603 & 140.69 & NA & NA\\
762 & 9.469866e+14 & 5.543771e+14 & 1.7827179 & 137.08 & 1.40e+18 & 1.00e+17\\
72 & 1.031184e+17 & 4.485403e+17 & 1.7669342 & 85.90 & 3.32e+18 & 8.49e+18\\
455 & 8.340879e+17 & 8.288418e+17 & 1.6642197 & 84.41 & 1.19e+18 & 1.20e+17\\
626 & 3.600162e+18 & 1.432526e+18 & 1.4705847 & 100.73 & 3.24e+18 & 1.30e+18\\
503 & 3.725514e+17 & 4.182849e+17 & 1.2278253 & 81.68 & 2.85e+18 & 3.40e+17\\
62 & 1.242051e+17 & 3.255760e+18 & 1.1981127 & 95.39 & NA & NA\\
97 & 1.830943e+18 & 1.484500e+18 & 1.1452324 & 82.83 & 1.33e+18 & 1.30e+17\\
675 & 2.053879e+18 & 1.689563e+18 & 0.8847463 & 76.00 & 1.20e+19 & 2.40e+18\\
618 & 1.354581e+15 & 7.772176e+14 & 0.8620717 & 120.29 & NA & NA\\
250 & 1.008119e+18 & 3.968719e+17 & 0.8082673 & 79.75 & NA & NA\\
287 & 9.492653e+17 & 2.409003e+17 & 0.7241231 & 67.60 & NA & NA\\
196 & 2.272704e+16 & 6.600523e+17 & 0.7131689 & 136.39 & 4.00e+18 & 1.53e+18\\
\hline
 \end{tabular}
\end{center}
\label{mass00}
\end{table}

\begin{table*}
\begin{center}
%\caption{Asteroid masses obtained with INPOP13c and compared with values extracted from \cite{CARRY2012}. This table gathers more specifically the asteroids inducing a maximum change of less than 5 meters on the Earth-Mars distances on a [1970:2015] interval.}
%\vspace{-0.4cm}
\begin{tabular}{c c c c c c c c c c }
\hline
IAU designation & INPOP13c & 1-$\sigma$ & Impact & diam & Carry 2013 & 1-$\sigma$\\
& kg & kg & m & km & kg & kg \\
33 & 3.321441e+18 & 1.068539e+18 & 0.6208345 & NA & 6.20e+18 & 7.40e+17\\
196 & 2.272704e+16 & 1.207070e+15 & 0.7131689 & 136.39 & 4.00e+18 & 1.53e+18 \\
779 & 1.311013e+18 & 3.501955e+17 & 0.5186778 & 76.62 & NA & NA\\
388 & 3.106367e+16 & 5.365290e+16 & 0.4414599 & 114.17 & NA & NA\\
204 & 2.385408e+16 & 8.976265e+16 & 0.2667205 & 48.57 & 6.00e+17 & 1.81e+18\\
\hline
  \end{tabular}
\end{center}
\label{mass00}
\end{table*}
%  \end{landscape}

\begin{table}
\caption{Values of dynamical parameters fitted to LLR observations. $GM_{EMB}$ is the sum of Earth's and Moon's masses, multiplied by the gravitationnal constant and is expressed in AU\textsuperscript{3}/day\textsuperscript{2}.
$C_{nmE}$ are the Earth's coefficients of potential (without unit).
$\tau_{21E}$ and $\tau_{22E}$ are time delays of the Earth used for tides effects and expressed in days.
$C_{nmM}$ and $S_{nmM}$ are the Moon's coefficients of potential (without unit). $(C/MR^2)_M$ is the ratio between the third moment of inertia of the Moon, divided by its mass and the square of the mean equatorial radius (without unit).
$k_{2M}$ and $\tau_M$ are the Love number (without unit) and the time delay (in day) of the Moon.
Formal errors at $1\sigma$ are given if the parameter is fitted and correspond to the values provided by the covariance matrix of the least square fit;
one can note that the real uncertainties on parameters are generally much higher.
Fixed values come from Lunar gravity model LP150Q \cite{2001Icar..150....1K} and Earth's ones from EGM96.}
\begin{center}
\begin{tabular}{c|r|l}
Name & \multicolumn{1}{c|}{Value} & \multicolumn{1}{c}{Formal error ($1\sigma$)} \\
\hline
$ GM_{EMB} $ 	 & $  8.9970115728 \times 10^{-10} $ & $ \pm 9.0 \times 10^{-19} $ \\
$ C_{20E} $ 	 & $ -1.0826225 \times 10^{-3} $ & $ \pm 1.0 \times 10^{-9} $ \\
$ C_{30E} $ 	 & $  2.756 \times 10^{-6} $ & $ \pm 3.7 \times 10^{-8} $ \\
$ C_{40E} $ 	 & $  1.6196215913670001 \times 10^{-6} $ &   \\
$ \tau_{21E} $ 	 & $  1.239 \times 10^{-2} $ & $ \pm 8.1 \times 10^{-5} $ \\
$ \tau_{22E} $ 	 & $  6.9768 \times 10^{-3} $ & $ \pm 6.7 \times 10^{-6} $ \\
$ C_{20M} $ 	 & $ -2.03377 \times 10^{-4} $ & $ \pm 1.8 \times 10^{-8} $ \\
$ C_{22M} $ 	 & $  2.23909 \times 10^{-5} $ & $ \pm 1.5 \times 10^{-9} $ \\
$ C_{30M} $ 	 & $ -8.4745310957091 \times 10^{-6} $ & \\
$ C_{31M} $ 	 & $  3.154 \times 10^{-5} $ & $ \pm 3.6 \times 10^{-7} $ \\
$ C_{32M} $ 	 & $  4.8452131769807101 \times 10^{-6} $ &   \\
$ C_{33M} $ 	 & $  1.7198 \times 10^{-6} $ & $ \pm 5.9 \times 10^{-9} $ \\
$ C_{40M} $ 	 & $  9.6422863508400007 \times 10^{-6} $ &   \\
$ C_{41M} $ 	 & $ -5.6926874002713197 \times 10^{-6} $ &   \\
$ C_{42M} $ 	 & $ -1.5861997682583101 \times 10^{-6} $ &   \\
$ C_{43M} $ 	 & $ -8.1204110561427604 \times 10^{-8} $ &   \\
$ C_{44M} $ 	 & $ -1.2739414703200301 \times 10^{-7} $ &   \\
$ S_{31M} $ 	 & $  3.203 \times 10^{-6} $ & $ \pm 7.3 \times 10^{-8} $ \\
$ S_{32M} $ 	 & $  1.68738 \times 10^{-6} $ & $ \pm 4.9 \times 10^{-10} $ \\
$ S_{33M} $ 	 & $ -2.4855254931699199 \times 10^{-7} $ &   \\
$ S_{41M} $ 	 & $  1.5743934836970999 \times 10^{-6} $ &   \\
$ S_{42M} $ 	 & $ -1.5173124037059000 \times 10^{-6} $ &   \\
$ S_{43M} $ 	 & $ -8.0279066452763596 \times 10^{-7} $ &   \\
$ S_{44M} $ 	 & $  8.3147478750240001 \times 10^{-8} $ &   \\
$ (C/MR^2)_{M} $ & $  3.93018 \times 10^{-1} $ & $ \pm 2.6 \times 10^{-5} $ \\
$ k_{2M} $ 	 & $  2.626 \times 10^{-2} $ & $ \pm 1.7 \times 10^{-4} $ \\
$ \tau_{M} $ 	 & $  1.912 \times 10^{-1} $ & $ \pm 1.2 \times 10^{-3} $ \\
\hline
\end{tabular}
\end{center}
\label{Tab_valeurs_parametres_LLR_dyn_I13b}
\end{table}

\begin{table}
\caption{Selenocentric coordinates of reflectors (expressed in meters) and Moon's Love number $h_2$ (without unit).}
\begin{center}
\begin{tabular}{cc|r|l}
Reflector & &\multicolumn{1}{c|}{Value} & \multicolumn{1}{c}{Formal error ($1\sigma$)} \\
\hline
           & x & $  1591924.511 $ & $ \pm 1.260 $ \\
Apollo XI  & y & $   690802.582 $ & $ \pm 2.890 $ \\
           & z & $    21003.774 $ & $ \pm 0.047 $ \\
\hline
           & x & $  1652725.840 $ & $ \pm 1.430 $ \\
Apollo XIV & y & $  -520890.307 $ & $ \pm 2.010 $ \\
           & z & $  -109730.480 $ & $ \pm 0.105 $ \\
\hline
           & x & $  1554674.570 $ & $ \pm 0.963 $ \\
Apollo XV  & y & $    98196.294 $ & $ \pm 3.000 $ \\
           & z & $   765005.696 $ & $ \pm 0.048 $ \\
\hline
           & x & $  1114345.496 $ & $ \pm 0.226 $ \\
Lunakhod 1 & y & $  -781226.597 $ & $ \pm 2.820 $ \\
           & z & $  1076059.335 $ & $ \pm 0.083 $ \\
\hline
           & x & $  1339314.148 $ & $ \pm 1.460 $ \\
Lunakhod 2 & y & $   801958.776 $ & $ \pm 2.430 $ \\
           & z & $   756359.229 $ & $ \pm 0.083 $ \\
\hline
Love number $h_2$ & & $ 5.26 \times 10^{-2} $ &  $\pm 2.4 \times 10^{-3} $   \\
\hline
\end{tabular}
\end{center}
\label{Tab_valeurs_parametres_LLR_I13b_reflecteurs}
\end{table}

\begin{table}
\caption{ITRF coordinates of stations at J1997.0, expressed in meters.}
\begin{center}
\begin{tabular}{cc|r|l}
Station & &\multicolumn{1}{c|}{Value} & \multicolumn{1}{c}{Formal error ($1\sigma$)} \\
\hline
                  & x & $  4581692.122 $ & $ \pm 0.003 $ \\
Cerga             & y & $   556196.023 $ & $ \pm 0.001 $ \\
                  & z & $  4389355.021 $ & $ \pm 0.010 $ \\
\hline
                  & x & $ -1330781.424 $ & $ \pm 0.011 $ \\
Mc Donald         & y & $ -5328755.458 $ & $ \pm 0.009 $ \\
                  & z & $  3235697.527 $ & $ \pm 0.021 $ \\
\hline
                  & x & $ -1330121.107 $ & $ \pm 0.013 $ \\
MLRS1             & y & $ -5328532.265 $ & $ \pm 0.008 $ \\
                  & z & $  3236146.619 $ & $ \pm 0.022 $ \\
\hline
                  & x & $ -1330021.436 $ & $ \pm 0.002 $ \\
MLRS2             & y & $ -5328403.283 $ & $ \pm 0.003 $ \\
                  & z & $  3236481.629 $ & $ \pm 0.010 $ \\
\hline
                  & x & $ -5466000.412 $ & $ \pm 0.011 $ \\
Haleakala (rec.)  & y & $ -2404424.718 $ & $ \pm 0.014 $ \\
                  & z & $  2242206.742 $ & $ \pm 0.028 $ \\
\hline
                  & x & $ -1463998.829 $ & $ \pm 0.003 $ \\
Apollo            & y & $ -5166632.678 $ & $ \pm 0.003 $ \\
                  & z & $  3435013.069 $ & $ \pm 0.010 $ \\
\hline
                  & x & $  4641978.858 $ & $ \pm 0.018 $ \\
Matera            & y & $  1393067.435 $ & $ \pm 0.027 $ \\
                  & z & $  4133249.634 $ & $ \pm 0.036 $ \\
\hline
\end{tabular}
\end{center}
\label{Tab_valeurs_parametres_LLR_I13b_stations}
\end{table}

\begin{table}
\caption{Means and standard deviations (both expressed in centimeters) of LLR residuals for INPOP13c solution. Na is the total number of observations available, Nk is the number kept in fitting process, Nr is the number that have been rejected according to the $3\sigma$ criterion (Na is always Nk+Nr).}
\begin{center}
\begin{tabular}{c|c|r|r|r|r|r}
Station & Period & Mean & Std. dev. & Na & Nk & Nr \\
\hline
Cerga      &  1984-1986  &   6.96 &  16.02 & 1188 &  1161 &   27 \\
Cerga      &  1987-1995  &  -0.58 &   6.58 & 3443 &  3411 &   32 \\
Cerga      &  1995-2006  &   0.06 &   3.97 & 4881 &  4845 &   36 \\
Cerga      &  2009-2013  &   0.08 &   6.08 &  999 &   990 &    9 \\
Mc Donald  &  1969-1983  &  -0.23 &  31.86 & 3410 &  3302 &  108 \\
Mc Donald  &  1983-1986  &   4.02 &  20.60 &  194 &   182 &   12 \\
MLRS1      &  1983-1984  &   5.92 &  29.43 &   44 &    44 &    0 \\
MLRS1      &  1984-1985  &  -7.25 &  77.25 &  368 &   358 &   10 \\
MLRS1      &  1985-1988  &   0.10 &   7.79 &  219 &   207 &   12 \\
MLRS2      &  1988-1996  &  -0.43 &   5.36 & 1199 &  1166 &   33 \\
MLRS2      &  1996-2012  &   0.15 &   5.81 & 2454 &  1972 &  482 \\
Haleakala  &  1984-1990  &  -0.38 &   8.63 &  770 &   739 &   31 \\
Apollo     &  2006-2010  &   0.25 &   4.92 &  941 &   940 &    1 \\
Apollo     &  2010-2012  &   0.00 &   6.61 &  514 &   414 &   10 \\
Apollo     &  2012-2013  &  -0.62 &   7.62 &  359 &   359 &    0 \\
Matera     &  2003-2013  &   0.13 &   7.05 &   83 &    70 &   13 \\
\end{tabular}
\end{center}
\label{Tab_residus_LLR_I13b}
\end{table}

\begin{table}
\caption{Comparison of residuals (in cm) between INPOP13c, DE421, DE430 and DE431. For the JPL's solutions, the residuals are obtained by applying  the same reduction model as INPOP13c to planetary and lunar motions from DExxx solutions, with all parameters of tables \ref{Tab_valeurs_parametres_LLR_I13b_reflecteurs} and \ref{Tab_valeurs_parametres_LLR_I13b_stations} refitted.}
\begin{center}
\begin{tabular}{c|c|r|r|r|r}
Station & Period & INPOP13c & DE421 & DE430 & DE431  \\
\hline
Cerga      &  1984-1986  &  16.02 &  14.56 & 14.55 &  14.19  \\
Cerga      &  1987-1995  &   6.58 &   5.87 &  5.64 &   6.27  \\
Cerga      &  1995-2006  &   3.97 &   3.97 &  3.96 &   4.91  \\
Cerga      &  2009-2013  &   6.08 &   3.84 &  3.52 &   6.94  \\
Mc Donald  &  1969-1983  &  31.86 &  30.38 & 30.15 &  30.43  \\
Mc Donald  &  1983-1986  &  20.60 &  19.54 & 19.83 &  19.58  \\
MLRS1      &  1983-1984  &  29.43 &  27.86 & 27.68 &  28.02  \\
MLRS1      &  1984-1985  &  77.25 &  73.80 & 72.70 &  73.85  \\
MLRS1      &  1985-1988  &   7.79 &   5.54 &  5.45 &   5.96  \\
MLRS2      &  1988-1996  &   5.36 &   5.16 &  4.88 &   5.49  \\
MLRS2      &  1996-2012  &   5.81 &   4.93 &  4.94 &   5.97  \\
Haleakala  &  1984-1990  &   8.63 &   8.59 &  8.33 &   8.76  \\
Apollo     &  2006-2010  &   4.92 &   4.47 &  4.15 &   6.02  \\
Apollo     &  2010-2012  &   6.61 &   4.99 &  4.31 &   7.85  \\
Apollo     &  2012-2013  &   7.62 &   5.09 &  4.43 &   8.95  \\
Matera     &  2003-2013  &   7.05 &   5.21 &  5.76 &   7.21  \\
\end{tabular}
\end{center}
\label{Tab_comp_INPOP_JPL}
\end{table}

\bibliography{biblio_hdr}{}
\bibliographystyle{plain}

\end{document}